\documentclass[final]{ias2}

\usepackage{graphicx} 
\usepackage{amsmath}
\usepackage{multirow}
\usepackage{array} 

\usepackage{hyperref} 

\begin{document}

\markboth{Pramana class file for \LaTeX 2e}{Jinfeng Liao}

\title{Anomalous transport effects and possible environmental symmetry ``violation''  
in heavy-ion collisions}

\author[iu,rbrc]{Jinfeng Liao} 
\email{liaoji@indiana.edu}
\address[iu]{Physics Department and Center for Exploration of Energy and Matter,
Indiana University, 
2401 N Milo B. Sampson Lane, Bloomington, IN 47408, USA.}
\address[rbrc]{RIKEN BNL Research Center, Brookhaven National Laboratory, 
   Upton, NY 11973, USA. }

\begin{abstract}
The heavy ion collision provides a unique many-body environment where local domains of strongly interacting chiral medium may occur and in a sense allow environmental symmetry ``violation'' phenomena. For example certain anomalous transport processes, forbidden in usual medium, become possible  in such domains. We briefly review recent progress in both the theoretical understanding and experimental search of various anomalous transport effects (such as the Chiral Magnetic Effect, Chiral Separation Effect, Chiral Electric Separation Effect, Chiral Electric/Magnetic Waves, etc) in the hot QCD fluid created by such collisions. 
\end{abstract}


 
\maketitle


\section{Introduction}

Exploring new states of matter under extreme conditions is a fundamental quest. It is believed that when heated to extremely high temperature the normal nuclear matter will change into a deconfined QCD matter known as the quark-gluon plasma (QGP). Such a quark-gluon plasma briefly occupied the Universe at about a few microseconds after the ``Big Bang'', and has now been (re-)created as the hottest matter ever in laboratory by heavy ion collisions (``Little Bang'')  at the Relativistic Heavy Ion Collider (RHIC) as well as at the Large Hadron Collider (LHC). To experimentally measure and theoretically understand the various properties of the QGP is  one of the major scientific frontiers in nuclear physics \cite{Jacobs:2007dw}. 

Successful experimental programs at both RHIC~\cite{RHIC_WP} and LHC~\cite{Muller:2012zq}, together with many theoretical and
phenomenological studies~\cite{review_GM,review_ES,Kolb:2003dz,Heinz:2009xj,Gyulassy:2003mc,Kovner:2003zj}, have
 revealed that the  created hot QCD   matter in heavy ion
collisions is a {\em strongly interacting}, nearly perfect fluid (thus so-called sQGP), opening avenues for probing highly non-trivial features of QCD. In particular, the sQGP provides a unique many-body environment for exploring a deep and fundamental aspect of QCD, namely anomalous effects arising from the interplay of topology, symmetry and anomaly in such matter and their possible experimental manifestations. In the past several years, a lot of progress has been made in this direction of study (see e.g. \cite{Kharzeev:2009fn,Kharzeev:2012ph,Kharzeev:2013ffa,Bzdak:2012ia}), which will be the focus of the present review.  The recent revival of interest in this topic has grown from a few deep origins of their own significance and interests, including the topological objects of field theories, the strong CP problem, and the axial anomaly of chiral fermions, which we will briefly discuss in the rest of this introduction section.  

The theoretical study of topological objects in field theories has a long history and has attracted persistent interests. Quite generically, such objects arise from embedding ``internal orientations'' (e.g. in color space) across real space-time boundaries in a topologically nontrivial manner, and are soliton solutions to the classical equations of motion 
for field theories due to the nonlinearity of the equations as well as
due to specific boundary conditions. 
They are found in field theories of various dimensions (1D kinks, 2D vortices, 3D monopoles, 4D instantons and sphalerons),  
and are known to be particularly important in the non-perturbative domain where the theories are strongly coupled (see e.g. \cite{'tHooft:1999au,Harvey:1996ur,Tong:2005un,Schafer:1996wv,ES_book}). Topological objects in Quantum Chromodynamics (QCD) are known to play important roles 
in many fundamental aspects such as the $U_A(1)$ anomaly~\cite{'tHooft:1999au}, the spontaneous chiral symmetry breaking a la the ``instanton liquid''~\cite{Schafer:1996wv,ES_book}, the confinement as dual superconductor of monopole condensate in the vacuum~\cite{Ripka:2003vv,Bali:1998de,Greensite:2003bk}, etc. More recently it has been found that these emergent objects are abundant and important not only for vacuum physics but also for the hot QGP at temperatures most relevant to RHIC and LHC~ \cite{Liao:2006ry,Liao:2011kr}. This may not be that surprising as close to the transition temperature $T \sim \,1-2 \, T_c \sim \Lambda_{QCD}$ the system is still quite strongly coupled  as also indicated by many lattice QCD studies \cite{D'Alessandro:2007su,Kaczmarek:2005ui,Bazavov:2012bq,Borsanyi:2011sw}). Let us give some more concrete discussions on the QCD instantons~\cite{Schafer:1996wv} and sphalerons~\cite{Shuryak_sphaleron,Moore} that are specifically relevant to the anomalous effects. These objects are processes connecting vacuum gluon field configurations with different Chern-Simons numbers, and accompanying all such processes is the characteristic topological winding number of these objects defined via their non-Abelian gauge fields as
\begin{eqnarray} \label{eq_Qw}
Q_w = \frac{g^2}{32\pi^2} \int d^4 x \,  F^{a}_{\mu \nu} \tilde{F}_{a}^{\mu \nu} 
\end{eqnarray}  
where $F_{\mu \nu}$ is the gauge field strength tensor and $\tilde{F}_{\mu \nu} = \frac{1}{2} \epsilon_{\mu\nu \rho\sigma} F^{\rho \sigma}$ the dual field strength tensor, with $a$ the color indices. 
The quantity $Q_w$ takes integer values by virtue of topological origin, e.g. $+1$ for one instanton and $-1$ for one anti-instanton. Let us emphasize one important feature: $Q_w \sim \int tr(\vec E\cdot \vec B)$ where $\vec E, \vec B$ are the corresponding chromo-electromagnetic fields. Recalling the symmetry properties of $\vec E$ and $\vec B$ fields, one realizes that $Q_w$ is P- and CP-odd and therefore certain P- and CP-odd effects may be induced by such topological objects. (One though also realizes that objects with opposite $Q_w$ occur with the same probability and on average have their effects canceling out, only allowing possible detection of certain fluctuations.) Note also for $Q_w$ being an integer being $\hat{o}(1)$, the field strengths in (\ref{eq_Qw}) must be $\vec E, \vec B\sim \hat{o}(1/g)$ which contribute $\vec E^2,\vec B^2 \sim \hat{o}(1/g^2)$ to the action cost for such topological objects, so they become important at strong coupling.  
With  such topological objects from deep theoretical origin  and with their effects deemed to be so important, 
a direct experimental manifestation of such objects or at least of certain unique imprints by them in high energy collider experiments, has been a long-sought goal of fundamental interest. 

Closely related to the topological objects is the strong CP problem~\cite{Quinn:2001vt,Peccei:1996ax,Kim:2008hd}. It is well-known that P and CP symmetries are broken in the electro-weak theory, and one may wonder if there is P and CP violation in QCD. Well, ultimately this is an experimental question. However, triggered by the study of topological objects in (\ref{eq_Qw}), it was  theoretically realized long ago that one may in principle add an additional $\theta$-term to the QCD Lagrangian without spoiling any other symmetries of QCD and the phenomenological success of QCD (except explicit P and CP violation) :  
\begin{eqnarray} \label{eq_theta}
{\cal L}_\theta =    \theta  \frac{g^2}{32\pi^2} F^{a}_{\mu \nu} \tilde{F}_{a}^{\mu \nu} 
\end{eqnarray}
Should such a term exists, there of course would be observable consequences, e.g. a nonzero neutron electric dipole moment (EDM). Dedicated high precision measurements of EDM so far suggest an upper limit $|\theta|<10^{-11}$. If one views $\theta$ as yet another ``god-given'' parameter of QCD, such smallish value may appear quite unnatural provided theoretically there is a prior no constraint on it. Alternatively one may think of the $\theta$-term as a low energy effective interaction by elevating $\theta$ into a dynamical field (known as ``axion'') that takes a vanishing expectation value at low energy due to certain interaction potentials in the high energy theory (e.g. with $<\theta>=0$ being the potential minimum). If this is indeed the case, then the value of $\theta$ becomes {\it environmental}. It could be quite plausible that in the hot QCD matter created in the high energy heavy ion collisions, the value of $\theta$ could fluctuate considerably from the vacuum one~\cite{Kharzeev:2009fn,Kharzeev:2013ffa,Kharzeev:2007tn,Zhitnitsky:2012im}. If there could be such environmental  P and CP symmetry violation, one may expect certain anomalous effects that become observable on an event-by-event basis in heavy ion collisions. This is such a fundamental aspect of QCD with far-reaching impacts on other physics branches (e.g. dark matter and cosmology~\cite{Kim:2008hd}), that it is certainly worth investing significant experimental efforts searching for its possible manifestation. 

Yet another  closely associated aspect is the axial anomaly when there are massless fermions present in the underlying theory and coupled to the gauge fields. Let us consider the vector and axial currents for one flavor of such fermion
\begin{eqnarray} \label{eq_currents}
J^\mu_V = \langle\bar{\psi} \gamma^\mu \psi\rangle  \quad , \quad  J_A^\mu=\langle\bar{\psi} \gamma^\mu\gamma_5 \psi\rangle
\end{eqnarray}  
(One can also introduce the left-handed(LH) and right-handed(RH) currents $J_{R/L} \equiv (J_V\pm J_A)/2$.) While both $J_V$ and $J_A$ are conserved classically, only $J_V$ is conserved at quantum level while the axial current conservation is spoiled by the axial anomaly:  
\begin{eqnarray} \label{eq_axial}
\partial_\mu J^\mu_A  =   \frac{g^2}{16\pi^2}   F^{a}_{\mu \nu} \tilde{F}_{a}^{\mu \nu} 
\end{eqnarray}
(For multi-flavor one simply scales with flavor number $N_f$.) One may notice that the same topological structure repeatedly occurs in (\ref{eq_Qw}),(\ref{eq_theta}), and (\ref{eq_axial}), indicating at deep connections among them. By integrating the anomaly relation over space-time volume, one obtains a very interesting result: 
\begin{eqnarray}
N_R - N_L = 2 \, Q_w \, ,
\end{eqnarray}
which implies that for processes involving a topological object with winding number $Q_w$, an imbalance of $2Q_w$ in the number of RH and LH fermion numbers is generated. Note that RH and LH fermions are parity mirrors of each other, and the generation of net chirality originates from the P and CP odd nature of topological objects as already pointed above. (Equivalently one may also rephrase similar generation of net chirality in the language of $\theta$-term via its time-dependence $\partial \theta/\partial t$, see discussions in e.g. \cite{Kharzeev:2009fn,Kharzeev:2013ffa}.) Now for QCD, it has two light flavors of quarks that in the vacuum ``suffer'' from spontaneous chiral symmetry breaking. However in the hot QGP phase with chiral symmetry restoration and with temperature scale much larger than the $u,d$ current quark masses,  one can consider them to be approximately massless, i.e. chiral. Therefore, if the hot QGP created in heavy ion collisions picks up nontrivial topological (or $\theta$) fluctuations, then net chirality of fermions would also be generated in this hot matter via axial anomaly. That is, one may expect certain local domain of matter with ``macroscopic'' net chirality --- a chiral medium possibly inducing P and CP odd phenomena.  

In short, studies from the past decade have shown that heavy ion collisions can create the quark-gluon plasma that flows in a nearly perfect manner well described by relativistic hydrodynamics. Such a QGP, with a temperature on the order of  the QCD nonperturbative scale $T\sim \Lambda_{QCD}$, is  strongly coupled with abundant topological structures emergent from non-Abelian gluon fields such as instantons and sphalerons: in this sense {\em the hot QCD fluid is a topological matter}.  With the possibility of  axion-type dynamics,   the $\theta$ value may fluctuate in hot dense environment away from zero: in this sense {\em the hot QCD fluid is a nontrivial $\theta$-matter}.   
With the temperature high enough to restore the chiral symmetry, the light flavor quarks in the QGP are approximately massless and net chirality can be generated via anomaly: in this sense {\em the hot QCD fluid is also a chiral matter}. 
A number of very interesting anomalous effects, that imply environmental symmetry ``violation'' and that otherwise would not occur, can arise in such a hot QCD fluid.  

The rest of the paper is organized as follows. In Section 2, we will discuss the various proposals of anomalous effects and the pertinent theoretical issues. Section 3 will focus on the relevant phenomenological studies and on the status of recent experimental measurements. Finally the summary and some concluding remarks will be given in the Section 4. Needless to say, this paper is by no means intended to be an inclusive collection of all relevant literature, which would be impossible for such a short overview. Rather it makes an attempt to outline some major theoretical ideas and to put an emphasis on the status of hunting for anomalous effects in heavy ion collisions. More detailed reviews with much fuller coverage of references can be found in e.g. \cite{Kharzeev:2012ph,Kharzeev:2013ffa,Bzdak:2012ia}.

\section{Anomalous effects in hot dense matter}

In this Section, we briefly introduce a number of theoretical ideas and some recent developments in understanding various anomalous effects in hot dense matter. Let us focus on one specific aspect of the transport properties, the matter's linear responses to externally applied Maxwell electromagnetic (EM) fields. Understanding such responses in matter is a fundamental quest with century-old history. The Ohm's law for conductors was known long ago, describing the generation of a vector current as a response to  electric field $\vec E$:  
 \begin{eqnarray} \label{eq_ohm}
\vec j_V= \sigma \vec E, \label{con}
\end{eqnarray}
where $\sigma$ is the electric conductivity of the matter with the convention that the electric current is $e \vec j_V$. Now in heavy ion collisions for the hot relativistic plasma with approximately chiral quarks, there are many more interesting possibilities to explore. First there are two possible currents, the vector and axial ones in (\ref{eq_currents}), that may be generated. Furthermore, the medium could be chiral (as discussed in details above),  i.e. with nonzero macroscopic chirality which may be characterized by a nonzero axial chemical potential $\mu_A$   in addition to usual temperature $T$ and vector chemical potential $\mu_V$. This therefore opens possibilities for interesting new transport processes and particularly anomalous ones in response to external EM fields. Lots of theoretical progress has been achieved, as we discuss below.

\subsection{The chiral magnetic effect}

The first and most famous example of such anomalous transport effects, is the chiral magnetic effect (CME), proposed and studied in the context of hot QCD matter first by Kharzeev~\cite{Kharzeev:2004ey}. A lot of works have been done to understand CME from various theoretical approaches \cite{Kharzeev:2007tn,Kharzeev:2007jp,Fukushima,Yee:2009vw,Buividovich:2009wi,Abramczyk:2009gb,Kalaydzhyan:2012ut}: again for a complete account of various developments see e.g. \cite{Kharzeev:2009fn,Kharzeev:2013ffa,Fukushima:2012vr}. The chiral magnet effect predicts the generation of a vector current along the direction of an external magnetic field $\vec B$:  
\begin{eqnarray}
\vec j_V=\sigma_5 \mu_A \vec B; \label{eq_cme}
\end{eqnarray}
with the chiral conductivity $ \sigma_5\equiv {N_ce \over 2\pi^2}$. This conductivity is a universal one with deep root in the chiral anomaly and has been found in both weak coupling and strong coupling calculations. 

It is not difficult to see the anomalous nature of the CME. The vector current is P-odd and CP-even while the magnetic field is P-even and CP-odd, so clearly generating $\vec j_V$ in parallel to $\vec B$ can not occur in normal circumstance: it can happen only in chiral medium with nonzero $\mu_A$ (which is a P- and CP-odd scalar quantity).  

To schematically understand the CME in a simple kinetic picture, imagine a plasma with more right-handed quarks than the left-handed ones --- meaning on average more quarks having momenta parallel to spins. The externally applied $\vec B$ (assuming up direction ) will align the quark spins, with the net result of more positive quarks spin-up and more negative quarks spin-down. Together with the net chirality it means there will be more positive quarks with momentum up and more negative quarks with momentum down: hence an electric current is induced in parallel to the $\vec B$.

\subsection{The chiral separation effect}

Another transport process also as response to the external $\vec B$ field and in close analogy to the CME, is the so-called  chiral separation effect (CSE) ~\cite{son:2004tq,Metlitski:2005pr}. The CSE predicts the generation of an axial current  along the  external $\vec B$ field at nonzero vector charge density (parameterized by the corresponding chemical potential $\mu_V$)  :
\begin{eqnarray}
\vec j_A=\sigma_5  \mu_V \vec B . \label{eq_cse}
\end{eqnarray} 
As it turns out, the same chiral conductivity $ \sigma_5 $ as in (\ref{eq_cme}) also appears in the above relation. This may not be accidental: the CSE and CME are like two manifestations of the same ``coin'' --- the axial anomaly. 

To schematically understand the CSE in a simple kinetic picture, imagine a plasma with more positive quarks than negative quarks (thus nonzero $\mu_V$). The externally applied $\vec B$ (assuming up direction) will align the quark spins, with the net result of more positive quarks spin-up and more negative quarks spin-down. Together with the imbalance between positive/negative quarks, it means the there will be more spin-up than spin-down quarks: so on average quarks with up-momentum carry positive helicity and quarks with down-momentum carry negative helicity, thus an axial current (with opposite helicity quarks moving in opposite directions) is induced  in parallel to the $\vec B$.

\subsection{The chiral electric separation effect}

Now if one reviews the Ohm's law (\ref{eq_ohm}), the CME (\ref{eq_cme}), and the CSE (\ref{eq_cse}) discussed above, they cover the generations of vector currents in response to $\vec E$ and  $\vec B$ and of axial current in response to $\vec B$, respectively. There is one entry in such transport properties that is missing: what about the generation of axial current in response to $\vec E$? 

 Recently this missing entry was filled by a novel effect found in ~\cite{Huang:2013iia}: the chiral electric separation effect (CESE). The CESE predicts the the generation of axial current in external $\vec E$ field when the medium has both nonzero vector and nonzero axial densities: 
 \begin{eqnarray}
\vec j_A = \chi_{ e} \mu_V \mu_A \vec E  \label{eq_cese} \, .
\end{eqnarray} 
We emphasize that in the above {\em a new type of transport coefficient $\chi_e$, the CESE conductivity}, has been introduced, which quantifies the medium's ability to generate axial current as a response to externally applied electric field.  Being a new transport property of a relativistic chiral plasma, it is of great theoretical interest to study $\chi_e$  for various systems and and in particular for the QGP. An explicit example is given in \cite{Huang:2013iia} for thermal QED plasma, with the computed CESE conductivity to be 
\begin{eqnarray}
\label{sigma_e}
\sigma_{\chi e}^{QED} =\chi_e \mu_V \mu_A  \approx 20.499\frac{\mu_V\mu_A}{T^2}\frac{T}{e^3\ln(1/e)} \,\, .
\end{eqnarray} 
More recently using the Hard-Thermal-Loop framework the CESE conductivity for the QGP is found to be 
\begin{eqnarray}
\label{sigma_e}
\sigma_{\chi e}^{QGP} = \chi_e \mu_V \mu_A  \approx (\#) T\frac{{\rm Tr}_{\rm f}[Q_eQ_A]}{g^4\ln(1/g)}  \frac{\mu_V \mu_A}{T^2}
\end{eqnarray} 
 to the leading-log accuracy with the numerical constant $(\#)$ depending on favor content, e.g. $(\#)=14.5163$ for $u,d$ light flavors \cite{Jiang:2014ura}.  Strong-coupling computation of this new conductivity has also been  obtained in certain holographic model~\cite{Pu:2014cwa} where it is found that $\sigma_{\chi e} \simeq T (\#) (8N_c\lambda/81)(\mu_V \mu_A/T^2)$ with the coefficient $\#\sim 0.025$. It is interesting to note that the weak-coupling and strong-coupling results share the same generic structural dependence $\sigma_{\chi e}\sim \mu_V \mu_A$ while have very different dependence on coupling constant --- the HTL result gives $\sigma_{\chi e} \sim 1/[g^4\ln(1/g)]$  while the holographic result gives $\sigma_{\chi e} \sim \lambda \sim g^2$.

It is also not difficult to see the anomalous nature of the CESE. While the axial current is P-even and CP-even,  the electric field is P-odd and CP-even and the vector density $\mu_V$ is P-even and CP-odd, so again generating $\vec j_A$ in parallel to $\vec E$ can  happen only in chiral medium with nonzero $\mu_A$. One however notices that under time reversal, the usual electric conductivity $\sigma$ and the CESE conductivity $\sigma_e$ are both T-odd (implying dissipation) which is different from the T-even chiral conductivity $\sigma_5$. Also unlike the $\sigma_5$, both the $\sigma$ and $\sigma_e$ are  specific properties of medium depending on the thermodynamic parameters and microscopic dynamics.

To schematically understand the CESE in a simple kinetic picture, imagine a conducting plasma with more positive quarks than negative quarks and with more right-handed quarks than left-handed ones. When an electric field is applied, the positive/negative quarks will move parallel/anti-parallel to the $\vec E$ direction. If there are more positive quarks  (that are moving along $\vec E$) and furthermore more right-handed ones, then the net result will  be a net flux of right-handed (positive) fermions moving parallel to $\vec E$. This picture is most transparent in the extreme situation when the system contains only right-handed quarks   (i.e. in the limit of $\mu_V=\mu_A>0$), with both a vector and an axial current concurrently generated in parallel to $\vec E$.

Finally, by putting the new CESE effect (\ref{eq_cese}) together with the Ohm (\ref{eq_ohm}), the CME (\ref{eq_cme}) and the CSE (\ref{eq_cse}), one   has the complete the table that characterizes a chiral medium's transport properties in response to externally applied EM fields: 
\begin{eqnarray}
\left(\begin{array}{c} \vec j_V \\ \vec j_A\end{array}\right) =  \left(\begin{array}{cc}
\sigma  & \sigma_5 \mu_A \\   \chi_e \mu_V \mu_A &  \sigma_5 \mu_V
\end{array}\right) \left(\begin{array}{c}
\vec E \\ \vec B
\end{array}\right). \label{eq_vaeb}
\end{eqnarray}

\subsection{Collective excitations and chiral electric/magnetic waves}

As can be seen in Eq.(\ref{eq_vaeb}), with the presence of external electromagnetic fields, the vector and axial densities/currents can mutually induce each other and get entangled together. In particular one may consider the small fluctuations of such densities, with the fluctuations of one generating the other and co-evolving along the external EM field direction. This is in analog to the pressure and energy density in hydrodynamics, with the fluctuation of one inducing the other and forming propagating collective modes i.e. the sound waves. Similarly here one can identify collective modes arising from the interplay between vector and axial densities via anomalous effects in EM fields. The first such example is the so-called Chiral Magnetic Wave found by Kharzeev and Yee in \cite{Kharzeev:2010gd}. Here let us follow the more general analysis in \cite{Huang:2013iia}. 

Let us consider a thermal   plasma in the static and homogeneous external $\vec E, \vec B$ fields, and study the coupled evolution of the very small fluctuations in vector and axial charge densities. Without loss of generality, we can always assume $\vec B$ is along $z$-axis, i.e., $\vec B = B \hat{z}$ while $\vec E =  E \hat{e}$. Using a linearized analysis of the coupled equations (\ref{eq_vaeb}), one can obtain a set of linear and coupled equations for the co-evolution of vector and axial density fluctuations (see details in  \cite{Huang:2013iia}). By Fourier-transforming the equations, one then can identify the dispersion relations for such collective modes with $(\omega,\vec k)$:  
\begin{eqnarray}
\label{eq_dispersion}
 \omega=-\frac{1}{2}\left[ ie\sigma_0-v_+(\hat{e}\cdot\vec k)\right]   \quad\pm
\frac{1}{2}\sqrt{\left[ ie\sigma_0-v_-(\hat{e}\cdot\vec k)\right]^2+4{\cal A}_\chi(\vec k)},
\end{eqnarray}
where $v_\pm=v_v\pm v_a$ with $v_v=2\sigma_2\alpha_V^2 n_V E$ and $v_a = \chi_e \alpha_V  \alpha_A n_V E$, and 
${\cal A}_\chi(\vec k) = \left[\sigma_5\alpha_A B(\hat{z}\cdot\vec k)+2\sigma_2\alpha_A^2 n_A E(\hat{e}\cdot\vec k)\right] \times\left[\sigma_5\alpha_V B(\hat{z}\cdot\vec k)+\chi_e\alpha_V\alpha_A n_A E(\hat{e}\cdot\vec k)\right]$. Here the $\sigma_0$, $\sigma_2$, $\chi_e$, $\alpha_{V}$, and $\alpha_A$ are various thermodynamic and transport coefficients, while $n_V$ and $n_A$ are background vector and axial densities (see  \cite{Huang:2013iia} for details). 

To see the physical meaning of the modes in (\ref{eq_dispersion}), let us consider two special cases. \\
(1) The case with only $\vec B = B \hat z$ field and $\vec E=0$. Then Eq.(\ref{eq_dispersion}) reduces to 
$\omega = \pm \sqrt{(v_\chi k_z)^2 - (e\sigma_0/2)^2 } - i (e\sigma_0/2)$ with speed $v_\chi = \sigma_5 \sqrt{\alpha_V \alpha_A} B$. When $v_\chi k_z \gg e\sigma_0/2$ we get two well-defined propagating modes 
\begin{eqnarray} \label{eq_cmw_dis}
\omega\approx  \pm v_\chi k_z -  i (e\sigma_0/2)  \, .
\end{eqnarray}
These are generalized {\em chiral magnetic waves (CMWs)} which reduce to the CMWs in \cite{Kharzeev:2010gd} when $\sigma_0=0$ and $\alpha_V=\alpha_A$. When $v_\chi k_z \le e\sigma_0/2$ the two modes become purely damped. \\
(2) The case with only $\vec E =E \hat z$ and $\vec B=0$. \\
First, we consider a background without vector density, i.e, $n_V=0$. In this case, we find two modes from (\ref{eq_dispersion}): 
$\omega = \pm \sqrt{(v_e k_z)^2 - (e\sigma_0/2)^2 } - i (e\sigma_0/2) $
with $v_e = \alpha_A n_A \sqrt{2\sigma_2\chi_e\alpha_V \alpha_A} E$. Similar to the CMWs, when $v_e k_z \gg e\sigma_0/2$ there are two  well-defined modes arising from CESE that propagate along $\vec E$ field:
\begin{eqnarray}
\omega\approx  \pm v_e k_z -  i (e\sigma_0/2)
\end{eqnarray}
 These two newly found modes can be called the {\em chiral electric waves (CEWs)}.  They become damped when $v_e k_z \le e\sigma_0/2$. \\
Second, if the background contains no axial density, i.e., $n_A=0$, then we see that the vector and axial modes become decoupled, and Eq. (\ref{eq_dispersion}) leads to
\begin{eqnarray}
\omega_V(\vec k)= v_v k_z - i (e\sigma_0)  \qquad 
\omega_A(\vec k)= v_a k_z.
\end{eqnarray}
The first solution $\omega_V(\vec k)$  represents a ``vector density wave" (VDW) with speed $v_v=2\sigma_2\alpha_V^2 n_V E$ that transports vector charges along $\vec E$ field but will be damped on a time scale $\sim 1/(e\sigma_0)$. The second solution $\omega_A(\vec k)$ is a new  mode arising from CESE and represents a propagating ``axial density wave" (ADW) along $\vec E$ with speed $v_a = \chi_e \alpha_V  \alpha_A n_V E$ and without damping.

In general when there are both $\vec E$ and $\vec B$ fields and both types of background densities, these physically distinct modes get mixed up into more complicated collective excitations in (\ref{eq_dispersion}). Clearly the external EM fields are the ``catalyst'', and the collective modes are the mutual induction and entangled propagation of the vector and axial density fluctuations. To see this nature of the collective modes in a more manifest way, let us take the chiral magnetic wave (in the simple case with $\sigma_0=0$ and $\alpha_V=\alpha_A=\alpha$) as an example. For the CMW in external $\vec B$ field, a fluctuation in the vector(axial) density $\mu_V$($\mu_A$) will induce axial(vector) three current $\vec j_A$($\vec j_V$) along $\vec B$ via CSE(CME) correspondingly, and such currents will transport the charge densities (as per continuity equations) along $\vec B$ which induce further currents. At the end there will be two waves propagating along $\vec B$ for the vector and axial charge densities respectively. These two waves can also be linearly recombined into two waves for the left-handed and right-handed densities via $j_{R/L}=j_V\pm j_A$. Mathematically the propagation of these waves can be described by the following wave equation~\cite{Kharzeev:2010gd}:
\begin{eqnarray} \label{eq_cmw}
\left(\partial_0 \mp{v_\chi} \partial_1 -D_L \partial^2_1 -D_T\partial^2_T\right) j^0_{L,R}=0, \label{cmw}
\end{eqnarray}
with $v_\chi=\frac{N_c e B \alpha}{2\pi^2}$ the velocity of the wave and $\alpha$ the susceptibility connecting charge density and chemical potential. The last two terms include also the dissipative effects due to diffusion of charge densities with $D_L$ ($D_T$) the longitudinal (transverse) diffusion constant with respect to the $\vec B$ direction. Such wave equations describe a right-handed density (a combination of vector and axial densities) fluctuation propagating in parallel direction to $\vec B$ while a left-handed density (another combination of vector and axial densities) fluctuation propagating in anti-parallel direction to $\vec B$.

\subsection{Discussions}

As already said, it is neither possible nor the purpose here to cover the various new developments on the anomalous effects. In this last part of theoretical discussions, let us just mention in passing a few very compelling ideas.

{\it Vorticity and kinetic helicity.---}It was realized a while ago~\cite{Kharzeev:2004ey,Kharzeev:2007tn,Son:2009tf,Kharzeev:2010gr} that external EM fields are not the only way to probe the anomalous responses. In fact one may utilize the so-called vorticity in hydrodynamics to play the role of a $\vec B$ field. To see that, think about a relativistic fluid with a four-velocity field $(u_0,\vec u)$ field --- this is just like the gauge vector field $(A_0,\vec A)$. Like $\vec B =\bigtriangledown \times \vec A$, the vorticity is $\vec \omega = \bigtriangledown \times \vec u$ and has similar symmetry properties to $\vec B$. One therefore may anticipate anomalous vector current induced by the vorticity in analog to the CME, $\vec j_V \propto \mu_A \mu_V \omega$, which has been called the chiral vortical effect (CVE). Since vorticity naturally occurs in the matter created in heavy ion collisions as a consequence of angular momentum conservation (see e.g.~\cite{Kharzeev:2004ey,Liang:2004ph,Becattini:2007sr}), it is of interest to explore the implications of CVE~\cite{Kharzeev:2010gr}. Recent STAR measurements reported at Quark Matter 2014 conference~\cite{STAR2014} appear to suggest a possible CVE-driven baryon separation along out-of-plane direction in analogy to the CME-driven charge separation. 
There are nontrivial implications if indeed CVE is present: for example CVE-driven separation effect could contribute substantially to the charge-separation as well and therefore the measured charge separation may be a combination of both CME and CVE contributions. One very useful way to help disentangle the two types of contribution, as pointed out recently, could be a further measurement of any possible {\it strangeness separation} along the out-of-plane direction, by using similar correlation observable for hadrons with nonzero strangeness charge. Lately it has been found that in analogy to the Chiral Magnetic Wave, there exists also vorticity-driven collective excitations, named as Chiral Vortical Wave (CVW) \cite{CVW_JHL}, that can induce highly nontrivial patterns for the splitting of elliptic flow between hadrons with opposite conserved charges.  
More generally speaking, in a relativistic chiral fluid with dynamical EM fields described by magnetohydrodynamics, there could be three types of ``chirality'' at play: the matter particles' net chirality, the kinetic helicity $\vec u \cdot \vec \omega$ (the handed-ness of the fluid field), and the EM field helicity $\vec A \cdot \vec B$ (the handed-ness of the EM fields) --- they can mutually transform and allow nontrivial anomalous transport processes that are yet to be fully investigated.   

{\it Anomalous hydrodynamics.---} While usually one considers the hydrodynamics to be the long wavelength effective theory  while the axial anomaly to be about the UV dynamics, people have found the delicate way for manifesting the anomaly at macroscopic level as anomalous terms in the hydrodynamic equations~\cite{Son:2009tf,Kharzeev:2011ds,Zakharov:2012vv}. To see that, while in usual hydrodynamics the current is simply $j^\mu=n u^\mu$, in anomalous hydrodynamics there will be such additional terms as $\sim D_B B^\mu$ (from CME) or  $\sim D \omega^\mu$ (from CVE), etc. Some of these coefficients could be fixed by general considerations like anomaly or entropy production. There are still many interesting aspects to explore within the anomalous hydrodynamics framework. To mention just two apparent examples: to manifest and study the CESE in this framework, and to explore the various collective modes in given hydrodynamic background and external EM fields. 

{\it Chiral kinetic theory.---} Another very interesting   developments concern the anomalous effects in the kinetic theory framework. Like hydrodynamics, the kinetic theory is another ``effective description'' at a different level of ``coarse-graining'' for describing many-body physics. How does the microscopic quantum anomaly manifest itself in kinetic evolution, and how do the anomalous effects emerge in the kinetic framework? There have been important progress in answering these questions~\cite{Son:2012wh,Stephanov:2012ki,Chen:2012ca,Loganayagam:2012pz} based on the Berry curvature in momentum space,  and effects like CME and CVE are understood in this framework. Many further works can be done in the chiral kinetic framework to  deepen our theoretical understanding of various known anomalous effects and recently found ones such as CESE. It may also be useful for help finding possible new anomalous effects and for determining certain anomalous transport coefficients that are not easily determinable otherwise.

\section{In search of the anomalous effects}

While all these anomalous effects bear their own theoretical significance and fundamental interest, there remains the big question: can one or more of such effects be experimentally observed in heavy ion collisions? This is extremely challenging as one is hunting for certain feeble traces in the  ``haystack'' of thousands of hadrons produced in each event. The evolution of the created matter from the initial conditions to the end products is very complicated and highly dynamical, involving several distinctive stages and with constituents strongly interacting for most of the time. One could foresee the rising of various possible ``background effects'' that may easily contaminate whatever signatures of the anomalous effects. Nevertheless different phenomenological studies and varied observables were proposed and a lot of analysis efforts have been made, with the search  still ongoing. In this short review, we focus on two sets of observables --- the charge-dependent azimuthal correlations and the charge-dependent elliptic flow, that have been thoroughly measured and have generated widespread interests to understand their precise meanings. Before that, we also give a discussion on recent progress in determining the strong EM fields in heavy ion collisions that are necessary for the anomalous effects to occur.

\subsection{The strong electromagnetic fields in heavy ion collisions}

Since the anomalous effects we discussed above (CME,CSE,CESE, and collective modes) arise only in the external EM fields and their magnitudes increase with the field strength, we need strong enough EM fields for such effects to actually occur and become detectable. As it turns out,  there are extremely strong transient $\vec E$ and $\vec B$ fields \cite{Kharzeev:2007jp,SIT_magnetic,Bzdak:2011yy,Deng:2012pc,Bloczynski:2012en,Tuchin:2010vs,Voronyuk:2011jd,McLerran:2013hla,Tuchin:2013apa} across the QCD plasma in heavy ion collisions, due to the electric charges (carried by the protons) in the heavy ions that are running at nearly the speed of light. In fact a simple order-of-magnitude estimate reveals that $|e\vec E|,|e\vec B|\sim \alpha_{EM} \, \gamma \, Z/ R_A^2 \sim m_\pi^2 \sim 10^{17} \rm Gauss$  (with $\gamma=\sqrt{s}/M_p$, $Z$ the number of protons in the nucleus, and $R_A$ the nuclear radius) is at the order of strong interaction scale and is   the strongest EM fields we know of in nature. This therefore provides the necessary strong external EM fields  for possibly generating the anomalous effects discussed above in the hot QGP created in heavy ion collisions. 

\begin{figure}
\begin{center}
\includegraphics[width=4.5cm]{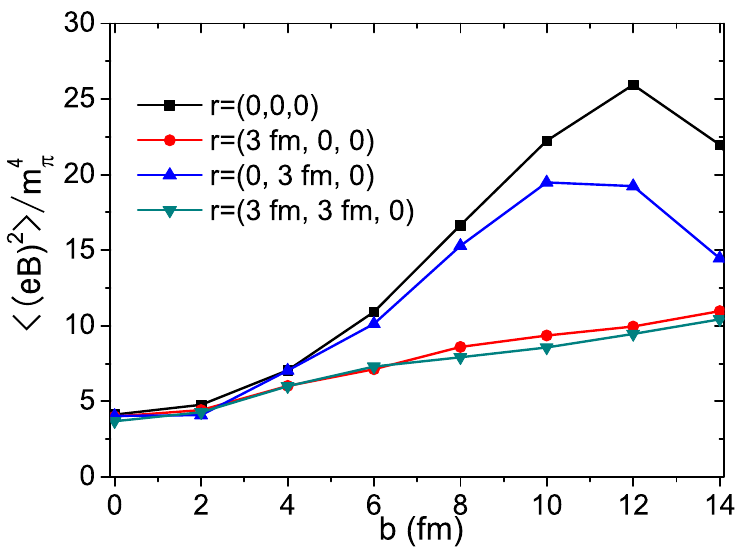} \hspace{1cm}
\includegraphics[width=4.5cm]{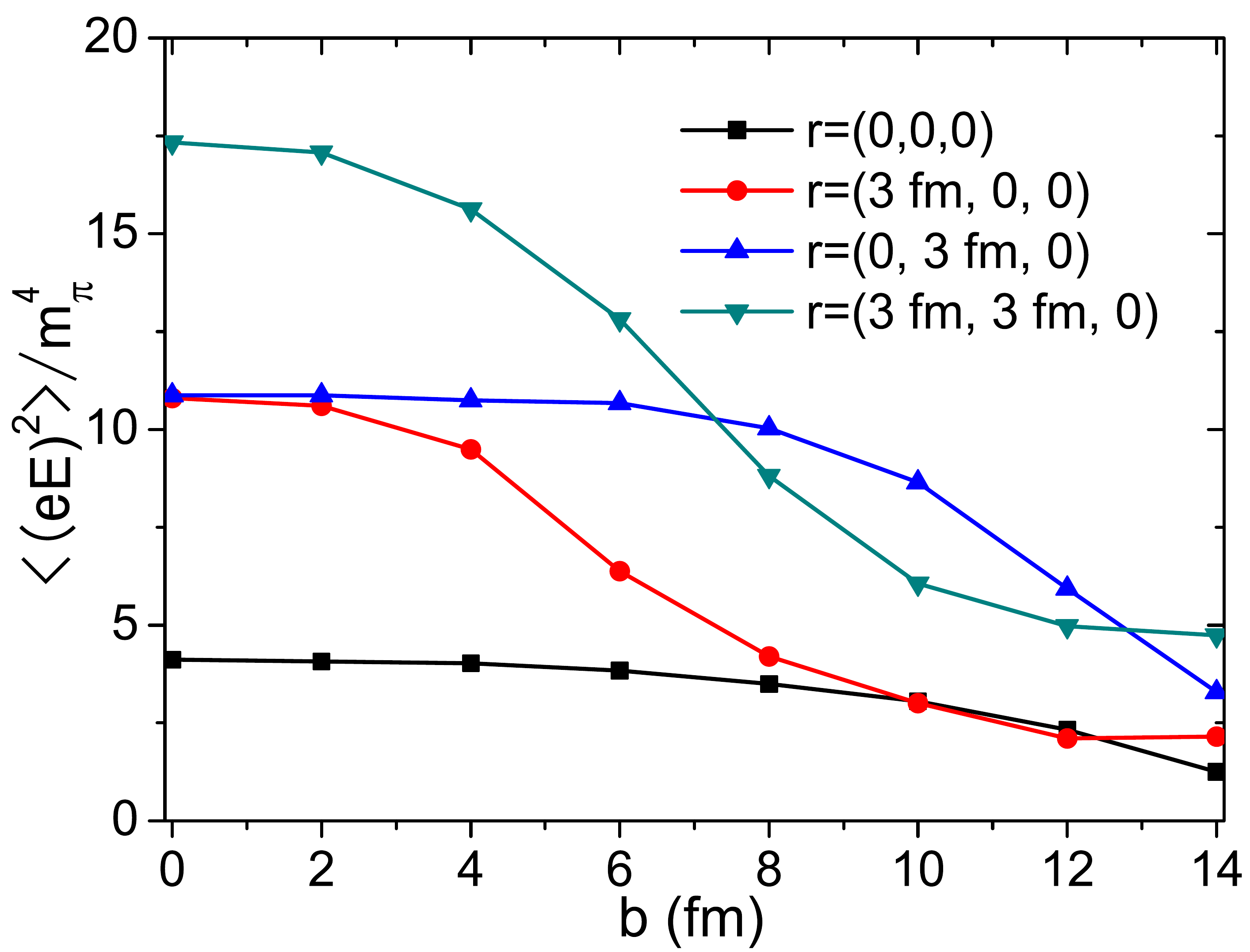}
\caption{(Color online) The event-averaged $(eB)^2$ and $(eE)^2$ (in unit of $m^4_\pi$) at $t=0$ and four different points on the
transverse plane
as functions of the impact parameter $b$ (from \cite{Bloczynski:2012en}).}
\label{fig_fields}
\end{center}
\vspace{-0.7cm}
\end{figure}

To precisely determine the strength of such EM fields is very important for quantitative phenomenology. Early computations of such fields were based on simple optical geometry of the colliding nuclei. In the past few years there were  major developments  in heavy ion physics showing the existence of strong initial state fluctuations~\cite{Alver:2010gr,Heinz:2013th}. Naturally such fluctuations would affect the EM fields derived from the protons whose initial positions inside nuclei fluctuate. A number of recent studies~\cite{Bzdak:2011yy,Deng:2012pc,Bloczynski:2012en} have  computed such EM fields with event-by-event initial fluctuations and found them to be  much enhanced in the central collisions as compared with the optical case. For example, we show the $\vec B$ and $\vec E$ strengths in Fig.\ref{fig_fields} (from \cite{Bloczynski:2012en}). One can see that even for perfectly central collision $b=0$ there are average fields of the strength $<(eB)^2> \approx 4 m_\pi^4$ while naive optical geometry would predict zero. Going to more peripheral collisions, the field strength first increases and then decreases, due to competition between increasing number of participants (which make dominant contribution to the fields) and increasing distance from the field point to spectator protons. For less central collisions, the impact of fluctuations is much less as the number of spectators becomes large and fluctuates less.

A decisive characteristic, essential for measurements in heavy ion collisions, is the {\em azimuthal orientation} of the $\vec B$ field. This is important to know, as effects like CME or CMW occur along  the $\vec B$ direction. From simple optical geometry consideration, it is not difficult to see that on average the $\vec B$ field points mostly in the out-of-plane direction (often labelled $\hat{y}$), i.e. in the direction perpendicular to the reaction plane (RP) spanned by the beam direction (often labelled $\hat{z}$) and the impact parameter $\vec b$ direction (often labelled $\hat{x}$). For a long time this has been taken for granted in all discussions of observable effects from such $\vec B$ field. This however is only partly true and an important step was recently made in \cite{Bloczynski:2012en}, showing for the first time that the sizable event-by-event fluctuations will spoil the azimuthal correlation between $\vec B$ and the matter geometry  to certain extent.

\begin{figure}[htbp]
\begin{center}
\includegraphics[width=4.cm]{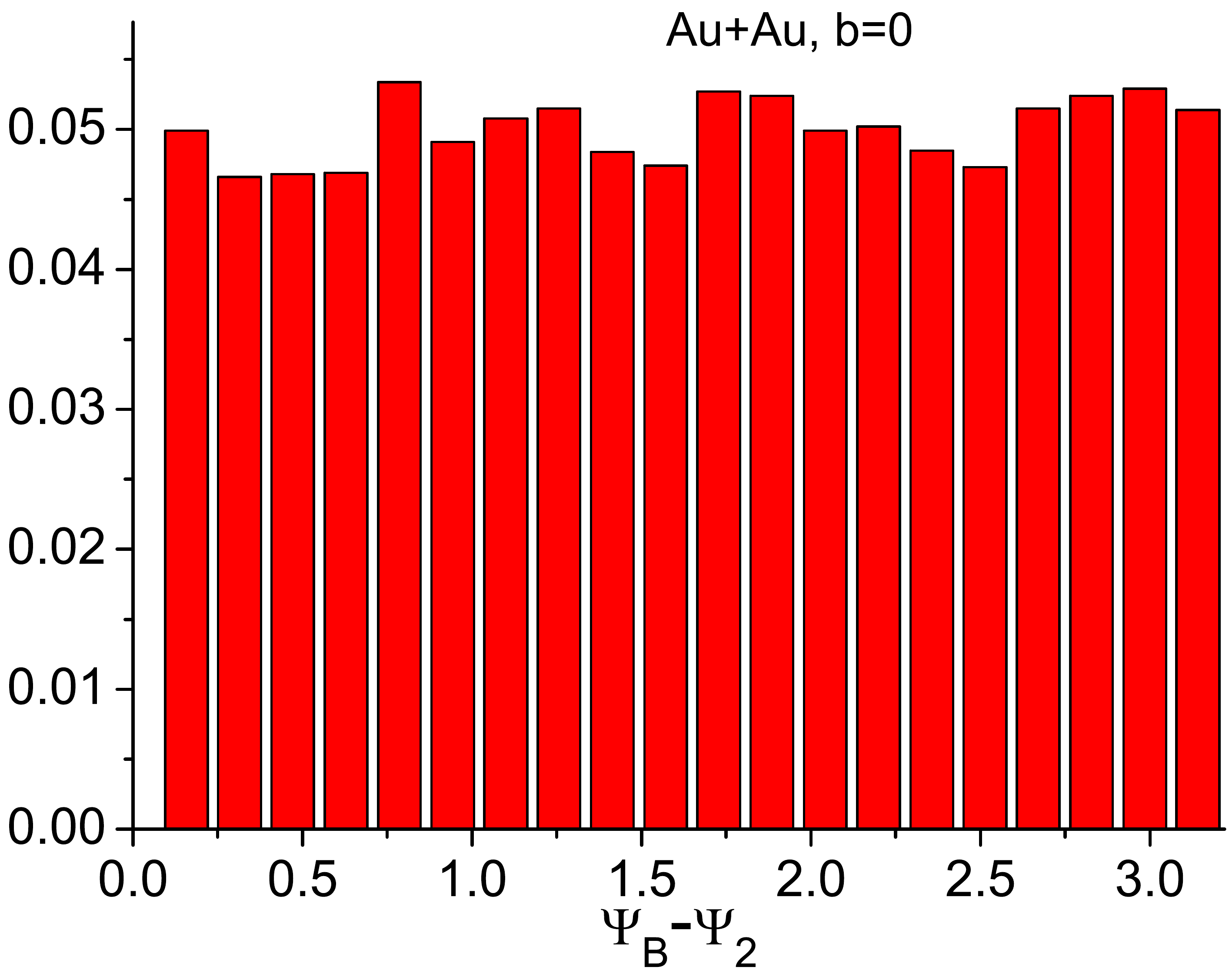}
\includegraphics[width=4.cm]{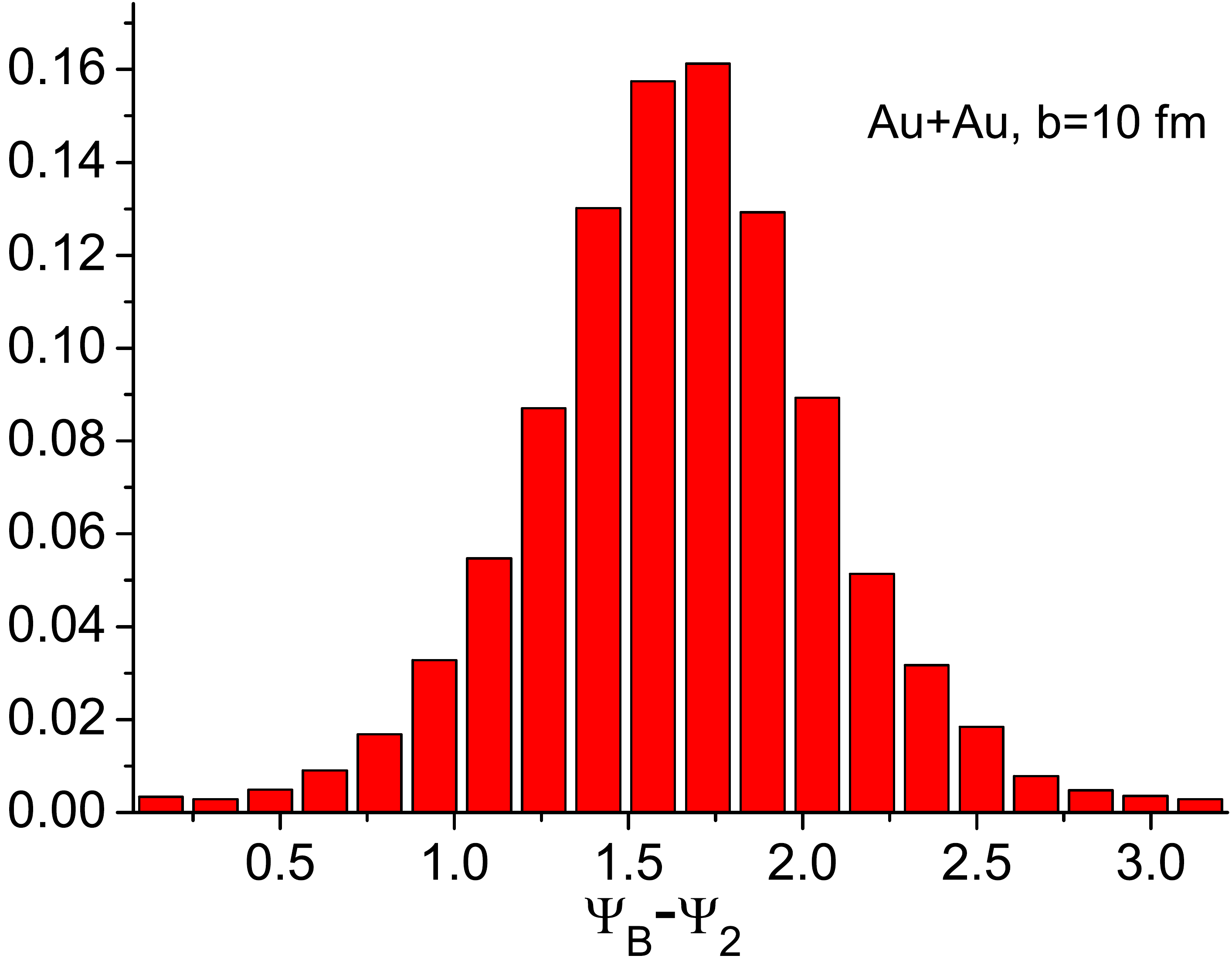}
\includegraphics[width=4.cm]{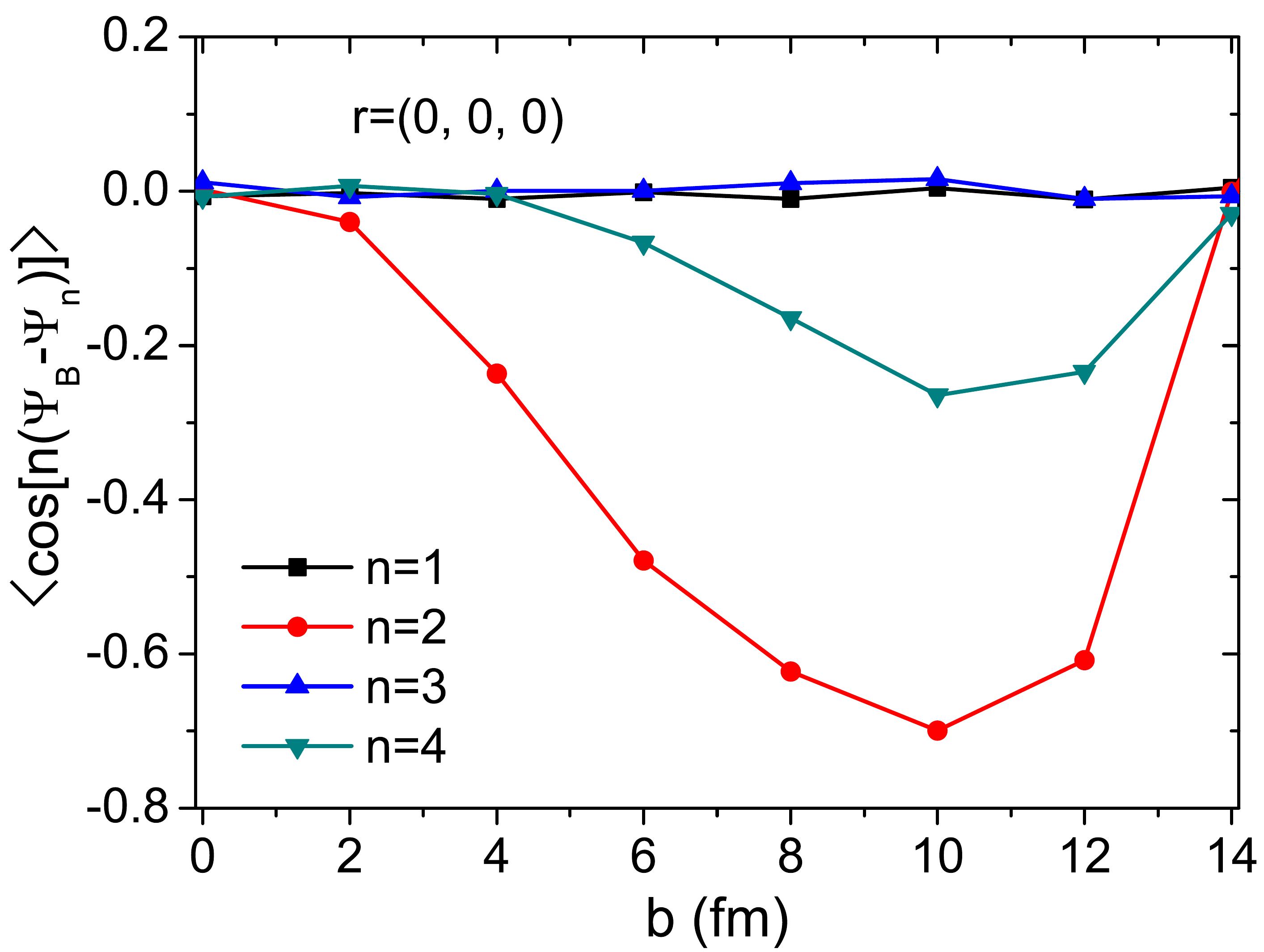}
\caption{(Color online) The event-by-event histograms of $\Psi_B-\Psi_2$ (in unit of radian) at impact parameters $b=0$fm (left) and $b=10$fm (middle) for Au + Au collisions at RHIC energy.  The azimuthal correlation factor $R_n$ in (\ref{eq_reduct1}) is also shown (right) as a function of $b$ for $n=1,2,3,4$.}
\label{fig_his}
\end{center}
\vspace{-0.2cm}
\end{figure}

Let us consider a heavy ion collision event. Due to fluctuations of nucleons' positions inside colliding nuclei, the initial matter's density distribution can be lumpy and irregular, and the azimuthal angle dependence can be decomposed into various harmonic ``participant planes'' characterized by certain angles $\Psi_n$ with respect to the ideal ``reaction plane''. Similarly in a given event the EM fields' transverse components $\vec B_\perp$ (and $\vec E_\perp$ as well) may also point toward some azimuthal direction $\Psi_{\bf B}$ other than that from the ``optical geometry''. What really matters for experimental measurements is the relative orientation of $\Psi_{\bf B}$ with respect to the matter geometry $\Psi_n$ in the very same event: clearly such relative orientation fluctuates from event to event. To see this, we show in Fig.\ref{fig_his} the event-by-event histograms of $\Psi_{\bf B}-\Psi_2$ at impact parameters $b=0$fm (left) and $b=10$fm (middle) for Au + Au collision at RHIC energy, where $\Psi_B$ is the azimuthal direction of $\vec B$ field (at $t=0$ and $\vec r=(0,0,0)$) and
$\Psi_2$ is the second harmonic participant plane.  Without fluctuations one strictly expects $\Psi_{\bf B} -\Psi_2=\pi/2$. However as one can see for the most central collisions $b=0$, the events are almost
uniformly distributed indicating negligible correlation between 
$\Psi_{\bf B}$ and $\Psi_2$. For $b=10$fm on the other hand, the event distributions evidently concentrate
around $\Psi_{\bf B} - \Psi_2= \pi/2 \approx 1.57$ indicating a tight correlation between the two. 
Such event-by-event fluctuations in the correlation between $\Psi_{\bf B}$ and $\Psi_n$  will typically bring in {\em a reduction to the measured signals from the intrinsic strength of any of the signals (of various field induced effects)} by the following factor~\cite{Bloczynski:2012en}:
\begin{eqnarray}
\label{eq_reduct1}
R_n=\langle\cos(n\bar{\Psi}^n_{\bf B})\rangle=\langle\cos[n(\Psi_{\bf B}-\Psi_n)]\rangle.
\end{eqnarray}
It is therefore necessary to study such correlations  and to quantify the above factor. The event-by-event computation results are shown in Fig.\ref{fig_his} (right):  $\Psi_{\bf B}$ are most strongly correlated with $\Psi_2$, visibly correlated with $\Psi_4$ while not correlated with $\Psi_{1,3}$. For the angular  correlation between $\Psi_{\bf B}$ and $\Psi_2$ (which in the optical limit would differ by $\pi/2$ with $R=-1$), it is smeared out significantly and vanishes in the very central and very peripheral collisions while stays strong for middle-centrality collisions. The apparent non-monotonic trend can be understood as follows: the $\Psi_{\bf B}$ is mostly determined by spectators whose number increases and fluctuates less with increasing $b$, while the matter geometry $\Psi_2$ is mostly determined by participants whose number whose number decreases and fluctuates more with increasing $b$, and hence there is the ``sweet-spot'' of $b\sim 10$fm where both have modest fluctuations and are most strongly correlated. 
This finding bears important 
 implications on observables related with various ${\bf B}$-induced effects (some of which will be discussed later) as pointed out  in \cite{Bloczynski:2012en}. It suggests that one ought not to expect any such effect to become detectable  in the very central collisions while the optimal centrality class for the search of these  effects corresponds to the impact parameter range of $b\sim8-12$fm.

Yet another important issue in full quantification of the strong EM fields, is their time evolution, which critically depends on the medium  feedback to the fast decaying fields \cite{McLerran:2013hla,Tuchin:2013apa}. If ignoring the possible medium feedback, the answer is simple: when the two colliding nuclei are approaching each other, the field strengths will rapidly increase toward the peak value upon collision and then decay rapidly with spectators/secondaries moving away from the collision zone~\cite{Deng:2012pc}. However if one consider the evolution of fields after collision to be affected by the dense partonic system created out of the collision, then it is more complicated. On general ground, with the $\vec B$ to rapidly decrease (without medium), a conducting medium will form induced-current that prevents the dropping of $\vec B$, i.e. a la the simple Lenz's law. The magnitude of this effect critically depends on the conductivity~\cite{Ding_conductivity,Gupta_conductivity} (thus the ability to create induced-current) of the underlying medium. There was the hope that the conductivity in the dense partonic matter might be high enough to bring the system into the magneto-hydrodynamic regime so the magnetic flux may be ``frozen'' with the fluid. However, recent analysis in \cite{McLerran:2013hla} appears to suggest that even with unrealistically large conductivity the magnetic field would still drop fairly quickly, thus one may not expect a lifetime of such fields to be beyond $\sim 1$fm/c time.  
  A conclusive answer may require more detailed understanding of the pre-thermal evolution~\cite{Blaizot:2011xf}, in particular the electric conductivity in such highly off-equilibrium partonic system. A closely related issue is how the field-induced effects should evolve with the collisional beam energy. This again is tricky. While the peak strength of $\vec B$ simply scales with Lorentz gamma factor $|eB|\sim \gamma = \sqrt{s} / M_p$ and thus grows quickly with beam energy, the time duration for the strong pulse scales inversely with the gamma factor $\tau \sim 1/\gamma$. As a result, if an effect is linear in field strength but accumulative in time (which is the case for most effects), then one may expect the total observable effect is tied to the time-integral of $\vec B$ field strength, thus becoming roughly independent of beam energy. This however may also be naive. With very high collision energy (e.g. at LHC), the field strength could become way stronger than medium scale and certain anomalous transport processes become nonlinearly dependent on $\vec B$ strength. With very low collision energy (e.g. at RHIC Beam Energy Scan), the QGP phase may become negligible or disappear altogether and the physics may become qualitatively different from that in the  hot QGP environment. These two issues, the time evolution and the beam energy dependence, are the important uncertainties to be better understood in future studies.

\subsection{The chiral magnetic effect and charge-dependent azimuthal correlations}

The Chiral Magnetic Effect (CME) \cite{Kharzeev:2007jp}  was the first proposed anomalous effect anticipated to be detectable in heavy ion collisions. Since in heavy ion collisions  
during the early stage  there are indeed very strong EM fields ($|e\vec B| \sim m_\pi^2$) and the $\vec B$ field is (approximately) pointing out-of-plane, one expects from the CME in (\ref{eq_cme}) that an electric current is generated in the out-of-plane direction and thus  induce a charge separation across the reaction plane i.e. more positive charges toward one pole of the fireball while more negative charges toward the opposite pole. Such a charge separation in coordinate space projects to  a charge dipole moment perpendicular to the reaction plane in the final hadron momentum space distribution, and should give rise to specific {\em charge-dependent azimuthal correlation patterns} that can be measured. One particular proposal ~\cite{Voloshin:2004vk} with sensitivity to a possible out-of-plane charge separation, is the following three-particle correlation
\mbox{$<\cos(\phi_i+\phi_j-2\phi_k)>$} for same-charge pairs
($i,j=++/--$) and opposite-charge pairs ($i,j=+-$) with the third
particle, denoted by index $k$, having any charge: 
\begin{eqnarray}
 <\cos(\phi_i+\phi_j-2\phi_k)>_{ij,\, k-any}\, \approx v_2 \, \gamma_{ij} \,\, , \nonumber \\  \gamma_{ij} \equiv <\cos(\phi_i+\phi_j-2\Psi_{RP})> \,\, . \quad \label{eq_gamma}
\end{eqnarray}
where the $\phi_{i,j,k}$ are azimuthal angles of the transverse momenta for these particles. An out-of-plane charge separation  would give a contribution $\gamma_{++/--}<0$ and $\gamma_{+-}>0$. 

\begin{figure}[htbp]
\begin{center}
\includegraphics[width=3.6cm]{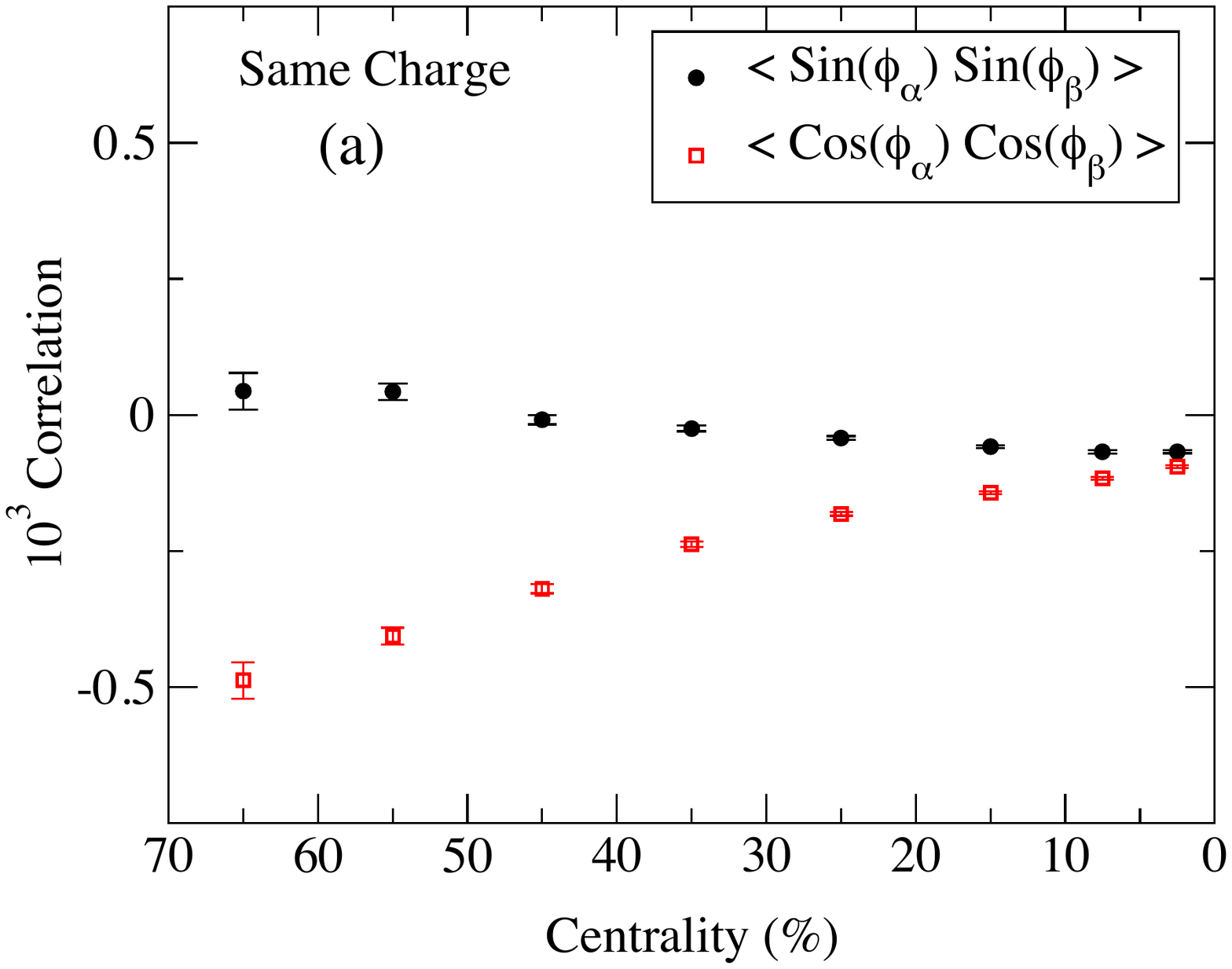}
\hspace{0.2cm}
\includegraphics[width=3.6cm]{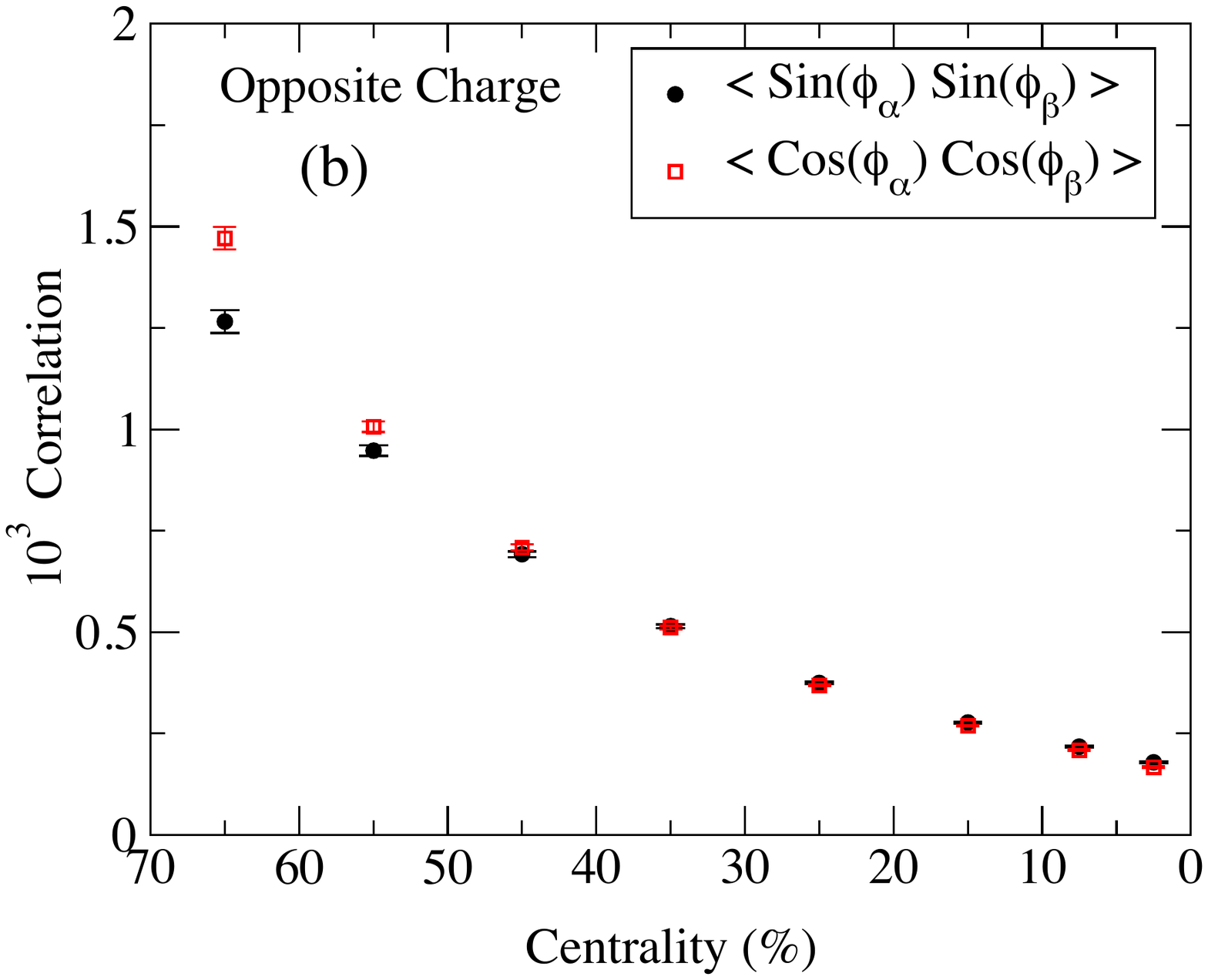}
\hspace{0.2cm}
\includegraphics[width=3.6cm]{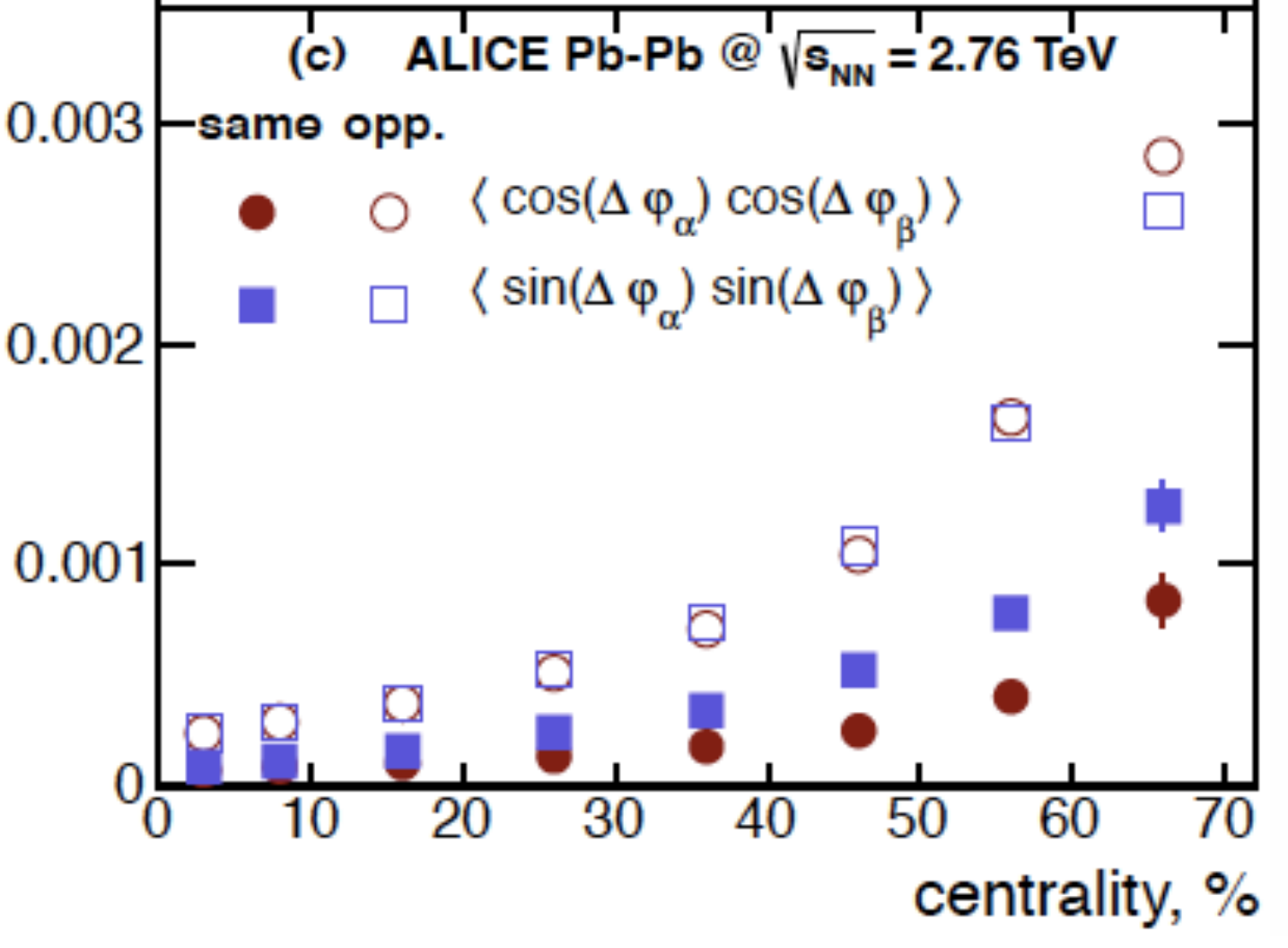}
\end{center}
\caption{The in-plane $\left\langle \cos (\protect\phi _{\protect%
\alpha })\cos (\protect\phi _{\protect\beta })\right\rangle $ and out-of-plane 
$\left\langle \sin (\protect\phi _{\protect\alpha })\sin (\protect\phi %
_{\protect\beta })\right\rangle $ projected correlations for same-charge (left) and opposite-charge (middle) pairs in RHIC $Au+Au$ collisions measured by STAR and in LHC $PbPb$ collisions (right) by ALICE.}
\label{fig_sscc}
\end{figure}

A few years ago, the STAR collaboration officially published data for the above correlations~\cite{Star:2009uh}, which showed a negative $\gamma_{++/--}$ correlation with the magnitude increasing from central to peripheral collisions. These  aspects are in line with CME expectations and the results generated strong initial enthusiasm in a wide community. 
Immediately following that, critical analysis of the precise implications of these data \cite{Bzdak:2009fc}, however,  showed features in data that significantly deviate from CME predictions. Specifically, for another two-particle correlation (also measured and published by STAR)
\begin{eqnarray}  \label{eq_delta}
\label{eqn_s_ij}  \delta_{ij}\equiv <\cos(\phi_i-\phi_j)>_{ij}\,\, ,
\end{eqnarray}
the CME prediction would be $\delta_{++/--}>0$ while the data showed quite the opposite, $\delta_{++/--} \approx \gamma_{++/--}<0$. In fact the analysis concluded that the dominant pattern in the data would be a back-to-back in-plane correlation for same charge pairs, obviously arising from some background effects. This situation is best revealed by the in-plane and out-of-plane projected correlations that can be converted from the $\gamma$ and $\delta$ correlators: 
\begin{eqnarray} 
<\cos\phi_i \cos \phi_j> =  (\delta_{ij} + \gamma_{ij}) /2  \, , \, <\sin\phi_i \sin \phi_j > =  (\delta_{ij} - \gamma_{ij}) /2  \quad
\end{eqnarray}
where for simplicity we set $\Psi_{RP}=0$. The projected correlations from STAR data are shown in Fig.\ref{fig_sscc} (left and middle). Measurements of these charge-dependent azimuthal  correlations at RHIC were also reported by PHENIX~\cite{Ajitanand:2010rc} later, showing quite similar features as the STAR data. The ``golden signature'' of CME could have been a positive out-of-plane same-charge correlation, i.e. $<\sin\phi_i \sin \phi_j >_{++/--} > 0$, while the RHIC data shows negligible amount of such correlation at almost all centralities. So, if CME is present, then there must also be other ``background'' effects canceling out the correlations from the CME contribution.  A series of subsequent studies~ \cite{Wang:2009kd,Pratt:2010gy,Bzdak:2010fd,Pratt,Hori:2012kp,Muller_separation,Muller_CME,Millo:2009ar} have identified and quantified a number of background effects, and also estimated the magnitude of possible CME signal. As of today, however, there is no firm conclusion yet on whether an unambiguous  CME signal would survive or not after background removal. Apart from the motivation of searching for CME, these correlations by themselves encode interesting dynamical information and present a challenge in understanding their origins. A very thorough discussion of these issues in the search for CME was given in a recent review article \cite{Bzdak:2012ia} and we refer the readers to that for details and references. In the rest of this subsection we will briefly highlight some important and more recent developments. 

\subsubsection{Discussions on background effects}

The initial idea behind the proposed correlator $\gamma_{ij} = <\cos\phi_i \cos\phi_j> - <\sin\phi_i \sin \phi_j>$ in (\ref{eq_gamma}) was that it measures a difference between in-plane versus out-of-plane correlations, and assuming backgrounds to give the same contributions in the two thus canceling out in the difference, it was hoped that the $\gamma$ correlator would just pick up the CME signal which is a pure out-of-plane signal $\sim \sin\phi_i \sin \phi_j$. The assumption about the backgrounds canceling out turns out to be wrong. A prominent phenomenon in heavy ion collision is the elliptic flow $v_2$, which basically tells that there is more  total out-of-plane momenta being carried by the expanding fluid than the total in-plane momenta. Even a background effect that is ``blind'' to reaction plane could pick up a difference in the contributions to the in-plane versus out-of-plane correlations simply due to the bulk elliptic flow. Let us show two examples here and also discuss possible way to disentangle them. 

One example is the transverse momentum conservation (TMC) ~\cite{Pratt:2010gy,Bzdak:2010fd}. To put it simple, since the total transverse momentum (i.e. the sum of all particles' momenta) of the fluid is zero and conserved over the whole evolution, the final observed hadrons must have their total transverse momentum to be zero. In a simple picture, imagine N particles each with the same momentum magnitude $P$ along the same axis, then an arbitrary particle's momentum is to be balanced by all the other particles with each balancing on average the amount of $\sim -P/(N-1)$: at large number of particles, this is simply $\sim - 1/N$ correlation in each direction. So the TMC will contribute a {\em negative} correlation $\sim -1 / N$ in both $ <\cos\phi_i \cos\phi_j> $ and $<\sin\phi_i \sin \phi_j>$ (thus stronger in more peripheral collisions with fewer particles), with  however the in-plane contribution slightly larger due to elliptic flow. Furthermore the TMC effect is {\em charge-independent}, i.e. contributing the same to the same-charge and opposite-charge pairs. Now looking at the Fig.\ref{fig_sscc}, it appears that there is a significant TMC contribution that is best manifested by the same-charge in-plane correlations (red symbol in left panel) while such a pattern is absent from other channels. Most interestingly, the same TMC contribution should apparently be present in the same-charge out-of-plane correlations while the actual data (black symbol in left panel) is clearly {\em positive} relative to the in-plane one (red symbol): this lends certain room to the possible scenario that the TMC contributes the common negative correlations in both in-plane and out-of-plane correlations, while the CME signal lifts the same-charge out-of-plane correlations to be more positive. 

The other example is the local charge conservation (LLC) ~\cite{Pratt:2010gy,Pratt}. Imagine the fluid cells (up to certain size) are perfectly charge neutralized, i.e being a charge canonical ensemble rather than grand-canonical. Upon freeze-out, the fluid cell will convert into a bunch of hadrons that first contain equal number of positive and negative charges and second on average co-move along the original momentum direction of the fluid motion. So a near-side opposite-charge pair correlation arises, with a few interesting features: the magnitude scales inversely with multiplicity, as only pairs from the same neutral cell are correlated and that will be diluted out by the total number of uncorrelated pairs;  the resulting correlation is slightly stronger in-plane than out-of-plane, again due to the elliptic flow; this correlation only contributes to the opposite-charge pairs. The LCC magnitude could be very strong if the radial flow is very large and the neutralized fluid cell size is large.  Again looking at the Fig.\ref{fig_sscc} (middle panel), it appears that indeed both the in-plane and out-of-plane correlations (with the former slightly stronger) for the opposite-charge pairs are well in line with the LCC expectations.    
 
Phenomenologically can one disentangle the flow-related and the CME-like contributions to the observed correlations? Attempts  were made recently \cite{Bzdak:2012ia,Bloczynski:2013mca}  based on a simple decomposition idea. Note that the CME-like signal only contributes the out-of-plane correlations, while the flow-related ones like the TMC and LCC contribute to both the in-plane and out-of-plane correlations with the former $\propto 1 + v_2$ and the latter $\propto 1-v_2$.  One can therefore make the following plausible two-component decomposition: $\gamma_{ij} = v_2 F_{ij} - H_{ij}$  and $\delta_{ij} = F_{ij} + H_{ij}$, 
where $F_{i j}$ are the $v_2$-related background contributions and $H_{ij}$ are the CME-like out-of-plane-only contributions. With the data for $\gamma$, $\delta$, and $v_2$ all available, one can extract the $F_{ij}$ and $H_{ij}$. Leaving out all the details, let us say that the remaining signal $H_{ij}$ does show a charge-dependence and the $H_{++/--}$ is positive and increasing toward peripheral collisions, matching the CME expectations. One though shall be cautious about the assumptions underlying such analysis. 
 
Clearly,  the current status cries out for very careful examinations and  for quantitative determination of the CME-induced signal. A realistic modeling incorporating these background effects together with CME in one simulation, would be highly desirable for a conclusive interpretation of the experimental data.

\subsubsection{Current status of measurements}

The charge-dependent correlations in (\ref{eq_gamma})(\ref{eq_delta}) have now been measured at a variety of collisional beam energies and also been analyzed in more differential ways. 

Toward the higher energy, ALICE collaboration has reported their measurements at LHC with beam energy $\sqrt{s}=2.76$TeV~\cite{Abelev:2012pa}. While the  $\gamma_{ij}$ correlations appear to be almost identical to that at RHIC, the $\delta_{ij}$ correlations change: in particular for the same-charge pairs the $\delta$ changes from being negative at RHIC to positive at LHC. As a consequence, when converted to the projected in-plane and out-of-plane correlations, as shown in Fig.\ref{fig_sscc} (right panel), for the first time one sees the desired positive signal for the same-charge out-of-plane correlations expected for the CME. One nevertheless has to be very careful here, and it is hard to draw any firm conclusion without a more solid theoretical/phenomenological understanding of how the CME and how the background effects should evolve from RHIC to LHC energies. There are a few other interesting features of these data as shown in Fig.\ref{fig_sscc} (right panel): 1) for the opposite-charge correlations, they follow similar patterns as at RHIC and become somewhat stronger, which might be understood through LCC with stronger radial flow at LHC; 2) for the same-charge in-plane correlations, it also becomes positive in contrast to the RHIC case where it is negative --- the positiveness at LHC in this channel clearly can not be attributed to just momentum conservation, so one has to worry about yet unknown source of such correlations. 

Toward the lower energy,  the STAR collaboration has reported measurements of the charge-dependent correlation (\ref{eq_gamma}) from the RHIC Beam Energy Scan \cite{Mohanty:2011nm,Wang:2012qs,Adamczyk:2014mzf}. The main message is that the difference between same-charge and opposite-charge pairs, i.e. the charge dependence, changes very little from top energy down to 19.6GeV while ceases to exist for collisions at 11.5GeV and lower. One may tend to interpret such trend as the ``turning off'' for one of the QGP signals, in light of the fact that a number of other key QGP signatures are gone in the same collision energy regime. This could be consistent with the CME expectation which relies upon a chiral-symmetric high-T QGP.  An important analysis made in ~\cite{Adamczyk:2014mzf} is to use experimentally measured $v_2$ for an attempt at decomposition of CME-driven signal and flow-driven background, based on an approach suggested in \cite{Bzdak:2012ia} and explored in \cite{Bloczynski:2013mca}. Very interestingly, the result suggests that given this approximate scheme of flow background subtraction, the remaining signal is consistent with CME predictions and gradually turns off with decreasing beam energy.   
Provided the complexity associated with the precise interpretation of such dynamical correlations already at the top energy collision, it will require future theoretical efforts that can quantitatively evaluate CME predictions as well as identified background effects within the same bulk evolution model in order to unequivocally evaluate the implications of these lower energy data for the search of CME. 

Another important new measurement was reported lately by STAR in \cite{Adamczyk:2013hsi}. Let us cut it short and simply mention the most interesting results from it (Fig.8 and 9 in the paper for 40-60\% centrality at 200GeV AuAu collisions). The projected out-of-plane correlation $<\sin\phi_i \sin \phi_j>$ was measured differentially in both the average pair rapidity $|<\eta>=(\eta_i+\eta_j)/2|$ and the pair relative rapidity $\Delta \eta = |\eta_i-\eta_j|$ as well as in both the average pair transverse momentum $<p_T>=(p_{T,i} + p_{T,j})/2$ and the pair relative transverse momentum $\Delta p_T = |p_{T,i}-p_{T,j}|$. The projected out-of-plane correlation for same-charge pair is found to be {\em positive} for: 1)  $|<\eta>| > 0.5$; 2) $\Delta \eta<0.2$; 3) $<p_T> < 0.3$GeV; and 4) $\Delta p_T < 0.1$GeV. That is, same-charge pairs with small average pair transverse momentum and relatively large rapidity yet with very small relative rapidity and very small relative transverse momentum appear to {\em co-move in the out-of-plane direction}. Qualitatively such a correlation in this particular kinematic regime is consistent with the  pattern predicted by the CME, which is interesting and encouraging. This though leaves the following challenging questions that remain to be understood: qualitatively why the CME-induced signal only survives in this kinematic regime while changes sign beyond, and quantitatively whether the CME prediction could give the correct magnitude as observed in the data.

\subsubsection{The Uranium-Uranium collisions}

As already discussed above, the current ambiguity in data interpretation is mostly related to elliptic flow $v_2$-related background effects. Since the $v_2$ is driven by initial geometry while the CME is driven by magnetic field,  it was proposed in ~\cite{Voloshin_UU} that a possible way of investigating those background effects is to measure the same correlations in the Uranium-Uranium (UU) collisions, due to the fact that the Uranium is a highly deformed (prolate) nucleus with a large quadrupole moment while the Au nucleus is nearly spherical. The interest in doing UU collisions has a long history with varied motivations~\cite{ES_UU,BaoAn_UU,Uli_UU,NFK_UU,Nu_UU} and such experiments were recently done at RHIC with some preliminary results reported in \cite{Wang:2012qs}. To see how the UU collisions may help, one realizes that when changing from the AuAu system to the UU system, the B field (that drives CME) and the initial matter geometry (e.g. eccentricity $\epsilon_2$ that drives elliptic flow) will change correspondingly in a quite distinctive way. For example, while in the most central collisions the $\vec B$ field will be very small for both systems, the UU will have a larger initial eccentricity (due to its prolate shape) as compared with the most central AuAu collisions. Such difference also persists at other centrality. One therefore expects that a possible CME signal and those flow-aided background effects would change differently from AuAu to UU, allowing a comparative study to possibly disentangle the two types of contributions. 

Recently a first attempt for an extrapolative study of these charge-dependent correlations' evolution from AuAu to UU was made in \cite{Bloczynski:2013mca}.  The  electromagnetic fields generated in the RHIC UU collisions
at $\sqrt{s}=193\, \rm GeV$ were systematically quantified, using event-by-event simulations incorporating initial state fluctuations. In particular the ``projected field strength''  $< (e {\bf B})^2\cos[2(\Psi_{\bf B}-\Psi_2)] >$  which controls the $\vec B$-field induced effects such as the CME was computed and compared with AuAu case. 
Taking the experimental data for charge-dependent azimuthal correlations from AuAu collisions,   a two-component decomposition (as discussed previously) was then introduced, based on  two types of identified sources ($v_2$-related and CME like) that could contribute to the measured correlations. One then further extrapolates  each component to UU system with reasonable assumptions: with $v_2$-related component scaling with $\epsilon_2$ while the CME-like component scaling with projected field strength. The resulting correlations were then compared with STAR data for UU collisions ~\cite{Wang:2012qs}. While there are clear discrepancy between the extrapolation results and the UU data at quantitative level, the extrapolation results are certainly in the right ballpark. The suggested two-component scenario with $v_2$-related and CME-like contributions might be a viable explanation of the measured charge-dependent azimuthal correlations, but this also calls for more careful modelings and more detailed experimental information.

\subsubsection{Possible future measurements}

We end by discussing a few proposals for future measurements that aim to help disentangle the background effects and clarify current understanding of the data. 

Since the CME predicted charge separation manifests as an out-of-plane charge dipole in the final hadrons' momentum space distribution, it may make sense to directly search for such a dipole. As suggested in  \cite{Liao:2010nv}, this can be achieved by the $\hat{Q}_1^c$ vector analysis (in analogue to the usual Q-vector analysis developed for measuring flow). In a given event of many measured hadrons,  the magnitude $Q^c_1$ and azimuthal angle $\Psi^c_1$ of
this vector can be determined in a given event by the following:
\begin{eqnarray} \label{eqn_qc1_def}
Q^c_1 \cos \Psi^c_1 \equiv \sum_i q_i \cos\phi_i  \,\,  , \,\,  Q^c_1 \sin \Psi^c_1 \equiv \sum_i q_i \sin\phi_i
\end{eqnarray}
where the summation is over all charged particles in the
event, with $q_i$ the electric charge and $\phi_i$  the azimuthal angle of each particle. In the same event the elliptic flow vector $\hat{Q}_2$ is determined similarly but in a charge-blind way. Examining the detailed distribution of the relative orientation between the $\hat{Q}_1^c$ and $\hat{Q}_2$ vectors over many events will help determine if there is an out-of-plane charge dipole. Provided current controversy in data interpretations and provided the feasibility of this analysis (which is no more complicated than the usual flow analysis), it seems such an analysis is definitely worth pursuing.

Another interesting proposal in \cite{Bzdak:2011np} attempts to utilize the differences in the fluctuations of eccentricity and EM fields to help separate the desired signal from backgrounds. It was found there that within the selected centrality class, if one examines the fluctuations of $\epsilon_2$ (which drives $v_2$ and $v_2$-related backgrounds) and the fluctuations of $\vec B$ field magnitude over many events, then one finds a significantly more fluctuations in $\epsilon_2$ than the $\vec B$ field. Therefore one may think of analysis similar to recently discussed ``event shape engineering'': for the many events in the given centrality class, one may look at the $\hat{Q}_2$ distribution and choose subsets of events with particularly small $v_2$ and particularly large $v_2$ (while with both subsets having similar strength of magnetic field), and then measure and compare the charge-dependent correlations $\gamma_{ij}$ in these two subsets. This could serve as a very useful analysis tool provided enough statistics.

\subsection{The chiral magnetic wave and charge-dependent elliptic flow}

More recently a new way of looking for subtle ``ripples'' from such anomalous effect was proposed, based on effects induced by   the collective excitations as discussed in Section 2.4. In particular, we focus on the chiral magnetic wave (CMW) that arises from the interplay between vector and axial charge densities through CME and CSE in an external $\vec B$ field. The dispersion of CMW is given by (\ref{eq_cmw_dis}) (ignoring the dissipation due to conductivity) while the wave equation is given by (\ref{eq_cmw}). 

\begin{figure}[htbp]
	\begin{center}
		\includegraphics[scale=0.65]{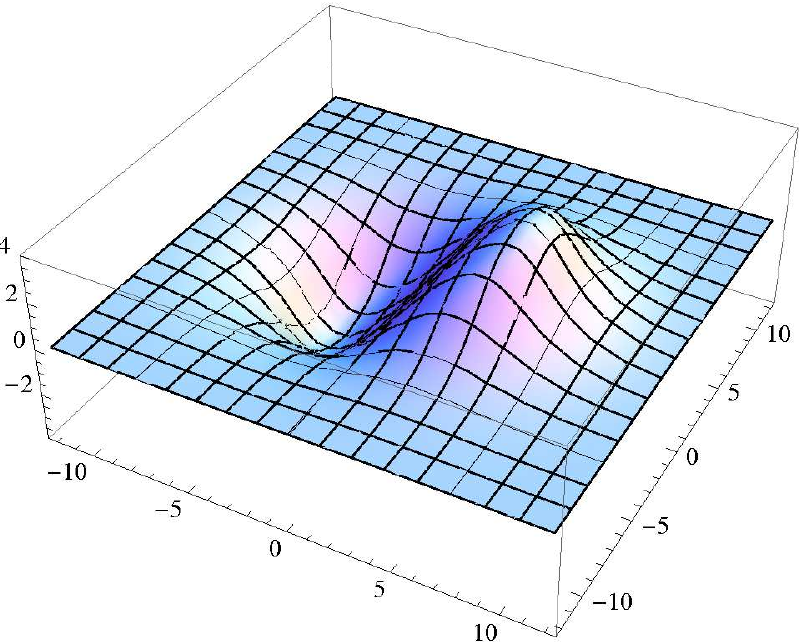} \hspace{0.3in}
			\includegraphics[scale=0.65]{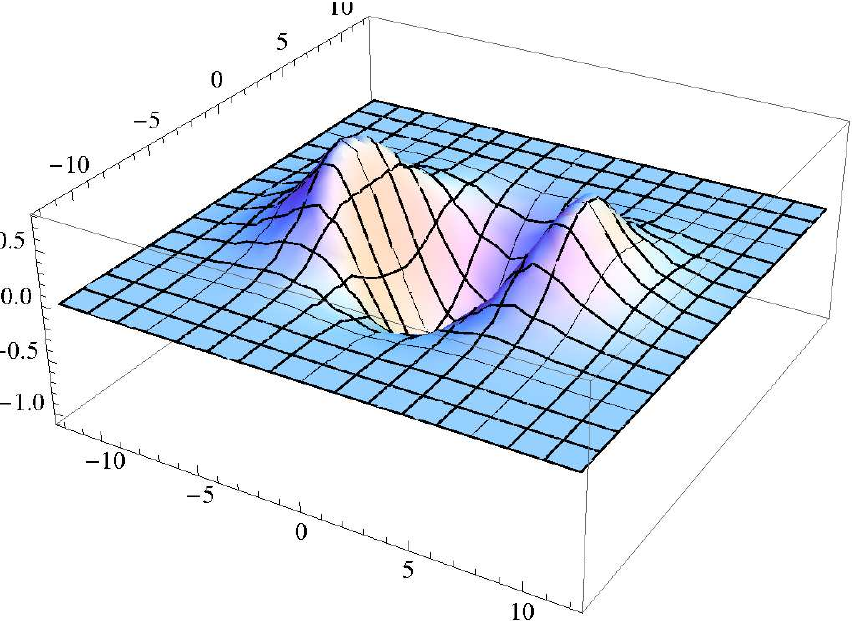} 
		\caption{ Axial charge density (left) and electric charge density (right) in the plane transverse to the beam axis (computed with magnetic field strength  $eB=m_\pi^2$, lifetime of magnetic field $\tau=10$ fm,  temperature $T=165$ MeV, impact parameter $b=3$ fm).}
		\label{chiral_chem_pot}
		\vspace{-0.5cm}
	\end{center}
\end{figure}

Imagine in a heavy ion collision the  QGP fireball is created with strong $\vec B$ field penetrating it at early time, what would be the possible consequence of chiral magnetic waves propagating in this medium? The question was first studied in  \cite{Burnier:2011bf}  and the conclusion is that such CMWs will induce an electric charge quadrupole moment (along the out-of-plane direction) for the created QGP. To schematically see how that happens: in the collision the overlapping zone picks up small nonzero vector charge density from the colliding nuclei; starting with such vector density in external $\vec B$ field (along the out-of-plane direction), the CMW will transport and separate positive/negative axial charge densities along $\vec B$, eventually leading to {\em a dipole moment of axial charge density}; the positive/negative axial densities at the two poles of the fireball will induce vector currents with opposite directions, transporting positive charges toward the poles while negative charges toward the equator of the fireball, eventually leading to  {\em a quadrupole moment of vector charge density (here the electric charge density)}; both the axial dipole and charge quadrupole are aligned along the $\vec B$ direction (see also ~\cite{Gorbar:2011ya}). Note that such initial vector charge density on average becomes larger and larger with decreasing beam energy, while even at high beam energy one can still have events with sizable initial vector density by fluctuations. By numerically solving the CMW equation (\ref{cmw}) with proper initial condition and  properties of QGP, it was shown in \cite{Burnier:2011bf} that the CMW evolution indeed leads to the axial charge dipole and electric charge quadrupole moments: see the plots in Fig.\ref{chiral_chem_pot}. 

\subsubsection{Phenomenological predictions and experimental data}

\begin{figure}[htbp]
\begin{center}
\includegraphics[scale=0.4]{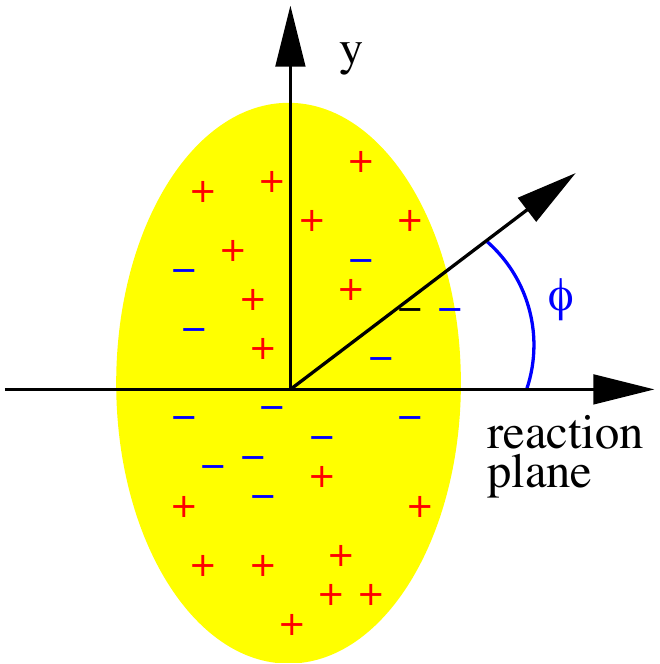}  \hspace{0.3cm}
\includegraphics[scale=0.33]{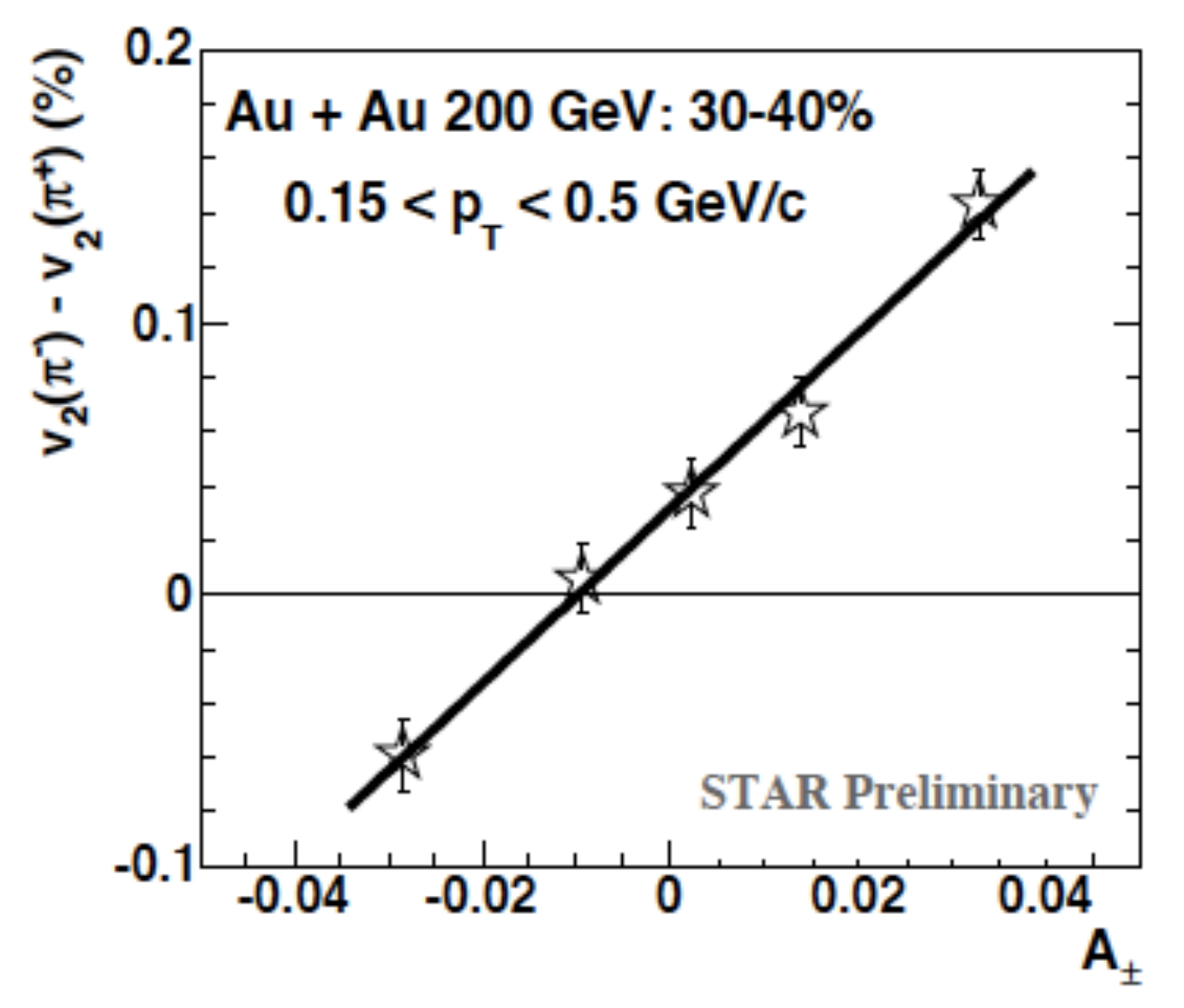}  \hspace{0.3cm}  
\includegraphics[scale=0.33]{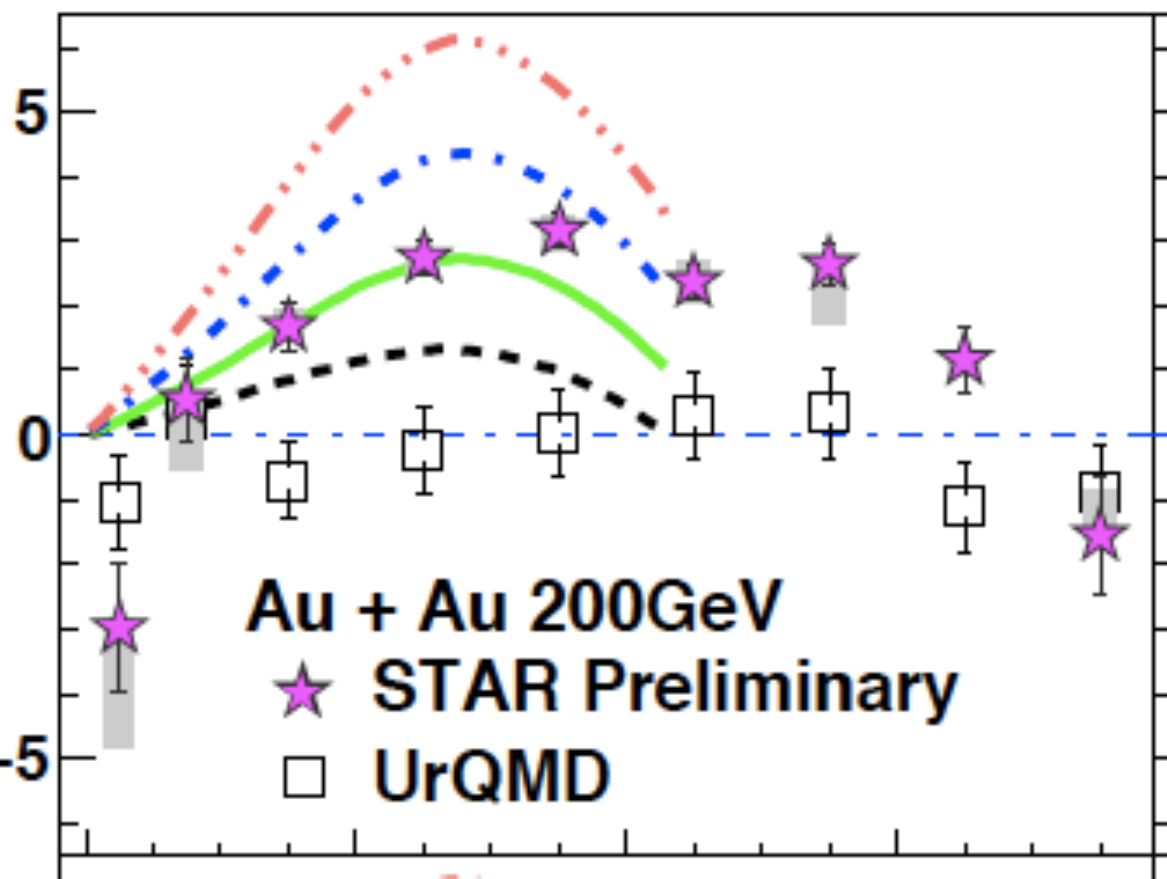}
\label{fig_v2}
\caption{ (left) Schematic demonstration of charge quadrupole being boosted by strong collective (radial) flow; (middle) elliptic flow splitting $v_2^{\pi^-}-v_2^{\pi^+}$ versus net charge asymmetry $A_{\pm}$ measured by STAR at $200\rm GeV$; (right) the slope parameter $r_e$(in \%) versus centrality, with black, green, blue, red lines from CMW calculations with magnetic field lifetime $\tau=3,4,5,6\, \rm fm/c$ respectively.}
\vspace{-0.5cm}
\end{center}
\end{figure}

Naturally one may wonder if the CMW induces a charge quadrupole for the QGP, how could that become measurable? It was proposed again in \cite{Burnier:2011bf} that such an electric charge quadrupole moment leads to a splitting between the positive/negative pions' elliptic flow. The idea is that such a spatial quadrupole charge distribution (at the end of the plasma phase) will be carried by strong radial flow and converted into azimuthal charge distribution in the final momentum space, resulting in more negative particles moving in-plane while more positive particles moving out-of-plane: see the demonstration in Fig.\ref{fig_v2} (left). To put it simple, the phenomenological prediction is a splitting between the $v_2$ of $\pi^\pm$ due to the CMW-induced electric quadrupole moment, which can be quantified by:
\begin{eqnarray}
v_2^{\pi^-} - v_2^{\pi^+}= r_e\,A_\pm \,\, . \label{eq_v2split}
\end{eqnarray}
where $A_\pm$ is the net charge asymmetry 
$A_\pm=\frac{\bar N_+-\bar N_-}{\bar N_+ +\bar N_-}$.  We see that there are three highly nontrivial aspects of such 
a prediction: (1) with positive $A_\pm$ (as is the case for lower beam energy collisions) the $\pi^-$ has a bigger elliptic flow than the $\pi^+$; (2) the splitting in $\pi^\pm$ elliptic flow is {\em linear} in charge asymmetry $A_\pm$; (3) the slope $r_e$ of such linear dependence is just  the quadrupole moment determined from net charge distribution due to the CMW evolution~\cite{Burnier:2011bf}. The size of the quadrupole $r_e$ was first computed in ~\cite{Burnier:2011bf} and  an improved calculation was presented in ~\cite{Burnier:2012ae}  with $\vec B$ field values from event-by-event simulations.  

Such a splitting was indeed first confirmed at low beam energies by STAR~\cite{Mohanty:2011nm}, in agreement with the prediction from \cite{Burnier:2011bf}. More recently the STAR collaboration has also systematically measured this $v_2$ difference as a function of net charge asymmetry at high collision energy~\cite{Wang:2012qs,Ke:2012qb}, which shows a {\em linear} dependence on $A_{\pm}$ just as predicted in (\ref{eq_v2split}): see the Fig.\ref{fig_v2} (middle panel). The magnitude of the extracted slope parameter $r_e$ and its centrality trend is also in good agreement with the CMW calculations~\cite{Burnier:2011bf,Burnier:2012ae}  assuming magnetic field lifetime $\tau=4 \, \rm fm/c $: see the Fig.\ref{fig_v2} (right panel). That is, all three nontrivial aspects of the CMW prediction in (\ref{eq_v2split}) appear to be confirmed in the experimental data with agreement at quantitative level, which is very encouraging.

\subsubsection{Anomalous hydrodynamic simulations}

With the  impressive experimental evidences in agreement with CMW prediction as discussed above,  it is crucial to advance the previous CMW computations (with certain simplifications and approximations)  toward a more realistic, comprehensive and quantitative simulation with outcome to be readily compared with measured data. Amongst other issues, two particularly important improvements are urgently needed: extending the CMW evolution equations to the case with an expanding background medium; incorporating a realistic description of the background medium evolution using the state-of-the-art hydrodynamic simulations that describe the bulk heavy ion data. 

A natural framework to implement such improvements is the anomalous hydrodynamics as briefly discussed in Section 2.5. Important progress has been made very recently by three different groups ~\cite{Hongo:2013cqa,Taghavi:2013ena,Yee:2013cya} in attempt to study the chiral magnetic wave using hydrodynamic background. The key question, of course, is quantitatively how much splitting could be generated from the CMW evolution in the more realistic anomalous hydrodynamic simulations in comparison with the experimental data. While with the common feature of implementing anomalous effects (CMW) into hydrodynamic evolution for RHIC fireball, these simulations have varied degrees of approximations and differ from each other in many details with some of them being important ones (e.g. equation of state, freeze-out procedure, hydro evolution, magnetic field, initial charge densities, etc). Not surprisingly, their conclusions are different from each other and consensus are not expected at the moment. Clearly these works represent first and important attempts toward a matured and mutually agreeable stage of anomalous hydrodynamic studies, and many efforts are needed for further improvements. Such a quantitative simulation of the CMW-induced contribution to the measured $v_2$-splitting is extremely crucial for a final conclusion, and we expect more progress to come along this line  in the near future.

\subsubsection{Discussions on possible subtleties}

While the agreement between CMW predictions and STAR measurements  is very encouraging thus far, one nevertheless needs to be very cautious about the precise interpretations of data. With all the complexities involved in the heavy ion collisions, there is certainly the possibility for one or more effects other than the CMW that may also contribute to the observed flow splitting pattern. There have already been a number of such proposals, e.g.  baryon stopping effect \cite{Dunlop}, mean-field potential at the hadronic stage \cite{Xu}, effect from net conserved charge current in hydro framework \cite{Steinheimer:2012bn}, charge-asymmetry-selection-bias effect in analysis \cite{Bzdak}, etc. It is though fair to say that so far none of these is as successful as the CMW effect particularly in light of the data at high beam energy. It will be crucial to have more detailed measurements for the dependence of such $v_2$ splitting on beam energies, on particle $p_t$ and $\eta$, as well as on particle identities (e.g. pions versus kaons), and possibly on any potential splitting of higher harmonic flows like $v_3$, which will all be useful  for constraining and distinguishing different models.

 A last point we'd like to discuss is an interesting feature in the splitting as shown in Fig.\ref{fig_v2} (middle panel): there is a {\em nonzero intercept} in the linear dependence of splitting on the charge asymmetry. That is, even with zero charge asymmetry there is a nonzero splitting (and in fact positive splitting) between the $\pi^\pm$ elliptic flow: this is a feature not in the original CMW predictions. It is of great interest to understand the origin of such an intercept. It was pointed out in \cite{Stephanov:2013tga} that two additional sources may contribute to such an intercept: first, the initial electric field $
 \vec E$ distribution on the transverse plane has a quadrupole structure and may contribute an additional charge quadrupole via simple $\vec E$-induced conducting current as in (\ref{eq_ohm}); second, the initial $\vec E\cdot \vec B$ distribution on the transverse plane has a dipole structure which may transform into a dipole in axial charge density and further induce additional contribution to charge quadrupole via the CME. Both are quite plausible ideas and appear to be supported by subsequent works in \cite{Hongo:2013cqa,Taghavi:2013ena,Yee:2013cya}.


\section{Summary}

In summary, we have given an overview of the theoretical ideas for anomalous effects in hot dense matter as well as the experimental search for such effects in relativistic heavy ion collisions. The hot QCD matter created in such collisions provides a unique system for exploring possible non-trivial environmental symmetry violation phenomena. A number of anomalous effects, such as the CME, CSE, CESE, and collective modes like CEW and CMW, have been theoretically found and studied in different frameworks. Specific observables for experimental search of such effects have been proposed and measured, including the charge-dependent azimuthal correlations for the CME and the charge-dependent elliptic flow for the CMW. We have discussed the current status of these measurements as well as our understanding and various issues in the data interpretations.  

There are other proposals for measuring anomalous effects and other strong-EM-field-induced effects in general that we have not covered~\cite{Hirono:2012rt,Basar,Yee:2013qma,Bali:2013owa,Shovkovy:2012zn,Chao:2013qpa,Gursoy:2014aka}. For example, it was suggested in ~\cite{Huang:2013iia} that the CESE in (\ref{eq_cese}) may induce specific charge azimuthal distribution pattern for created QGP in the asymmetric Cu-Au collisions at RHIC where there is strong in-plane electric field pointing from Au to Cu \cite{Hirono:2012rt}. Such electric field in Cu-Au collisions may also induce simple conducting current leading to measurable in-plane charge dipole that may be measurable~\cite{Hirono:2012rt}. There was also proposal that such strong EM fields could induce photon emissions with large azimuthal anisotropy that could explain data~\cite{Basar,Yee:2013qma}. These studies are valuable toward a coherent understanding of various effects arising from the common EM fields in heavy ion collisions.   

There are a number of particular challenges that need to be addressed. The time evolution of EM fields need to be better understood and implemented into various phenomenological modelings. Currently it seems the evolutions of either the charge-dependent correlations or the charge-dependent elliptic flow with collisional beam energy could provide very useful constraints on different interpretations of data, and more investigations are needed to understand such evolutions with beam energy. All above all, more realistic and comprehensive simulations of CME, CMW and other background effects will be indispensable for either an approval of a disapproval eventually for the search of anomalous effects in heavy ion collisions. 

Let us stop here by just emphasizing again the great opportunity provided by the hot dense matter created in heavy ion collisions: it allows exploring effects arising from fundamental aspects like topology and anomaly of quantum field theories, and allows investigating whether fundamental symmetries like P and CP in QCD could become environmental by controlled and repeatable experiments in laboratories. Needless to say, such a search is extremely challenging and may become frustrating. But one may recall the years it took to find the Higgs boson eventually, and the still ongoing search for Dirac's 1931 proposal of magnetic monopoles in electrodynamics. With the deep and fundamental implications that the search of anomalous effects may bring us, it is a goal well worth more joint and dedicated efforts of theoretical, phenomenological, and experimental studies.

\acknowledgments
The author thanks J. Bloczynski, Y. Burnier, A. Bzdak, X. Huang, Y. Jiang, D. Kharzeev, V. Koch, H. Yee, and X. Zhang for collaborations on various related aspects. He also thanks H. Huang, M. Huang, H. Ke, R. Lacey, L. McLerran, B. Mohanty, M. Stephanov, A. Tang, G. Wang, and N. Xu for discussions and communications. The research of the author is supported by the National Science Foundation (Grant No. PHY-1352368) and he is also grateful to the RIKEN BNL Research Center for partial support. The author thanks the Yukawa Institute for Theoretical Physics, Kyoto University, where this work was partly completed during the YITP-T-13-05 on ``New Frontiers in QCD''. 

\bibliographystyle{pramana}
\bibliography{references}

\end{document}